\documentclass[lettersize,journal]{IEEEtran}
\usepackage{amsmath,amsfonts}
\usepackage{algorithmic}
\usepackage{algorithm}
\usepackage{array}
\usepackage[caption=false,font=normalsize,labelfont=sf,textfont=sf]{subfig}
\usepackage{textcomp}
\usepackage{stfloats}
\usepackage{url}
\usepackage{verbatim}
\usepackage{graphicx}
\usepackage{cite}
\usepackage{hyperref}
\usepackage{color}
\usepackage{amsthm}
\usepackage{listings}
\usepackage{siunitx}
\usepackage{bbm}
\usepackage{numprint}
\npthousandsep{,}\npthousandthpartsep{}\npdecimalsign{.}
\usepackage{dsfont}
\usepackage{multirow}

\newtheorem{proposition}{Proposition}
\newtheorem{remark}{Remark}

\newtheorem{corollary}{Corollary}

\definecolor{codegreen}{rgb}{0,0.6,0}
\definecolor{codegray}{rgb}{0.5,0.5,0.5}
\definecolor{codepurple}{rgb}{0.58,0,0.82}
\definecolor{backcolour}{rgb}{0.95,0.95,0.92}

\lstdefinestyle{mystyle}{
    backgroundcolor=\color{white},   
    commentstyle=\color{codegreen},
    keywordstyle=\color{magenta},
    numberstyle=\tiny\color{codegray},
    stringstyle=\color{codepurple},
    basicstyle=\ttfamily\footnotesize,
    breakatwhitespace=false,         
    breaklines=true,                 
    captionpos=b,                    
    keepspaces=true,                 
    numbers=left,                    
    numbersep=5pt,                  
    showspaces=false,                
    showstringspaces=false,
    showtabs=false,                  
    tabsize=2
}

\newcommand{\ldn}[1]{{\color{blue}#1}}

\newcommand{\new}[1]{\textcolor{black}{#1}}\newcommand{\old}[1]{\iffalse {#1} \fi}

\newcommand{\newre}[1]{\textcolor{black}{#1}}\newcommand{\oldre}[1]{\iffalse {#1} \fi}

\begin{document}

\title{SIMBa: System Identification Methods leveraging Backpropagation}

\author{Loris Di Natale, Muhammad Zakwan, Philipp Heer, Giancarlo Ferrari-Trecate, Colin Neil Jones
\thanks{This research was supported by the Swiss National Science Foundation under NCCR Automation, grant agreement 51NF40\_180545. \textit{(Loris Di Natale and Muhammad Zakwan contributed equally to this work.)} (\textit{Corresponding author: Loris Di Natale}.)}
\thanks{Loris Di Natale, Muhammad Zakwan, Giancarlo Ferrari-Trecate, and Colin Neil Jones are with the Laboratoire d'Automatique, Swiss Federal Institute of Technology Lausanne (EPFL), 1015 Lausanne, Switzerland (e-mail: loris.dinatale@alumni.epfl.ch; muhammad.zakwan@epfl.ch; giancarlo.ferraritrecate@epfl.ch; colin.jones@epfl.ch).}%
\thanks{Loris Di Natale and Philipp Heer are with the Urban Energy Systems Laboratory, Swiss Federal Laboratories for Materials Science and Technology (Empa), 8600 D\"{u}bendorf, Switzerland (e-mail: philipp.heer@empa.ch).}%
}

\markboth{Preprint}
{Shell \MakeLowercase{\textit{et al.}}: A Sample Article Using IEEEtran.cls for IEEE Journals}


\maketitle

\begin{abstract}

This manuscript details \new{and extends} the SIMBa toolbox (System Identification Methods leveraging Backpropagation) \new{presented in previous work}, which uses well-established Machine Learning tools for discrete-time linear multi-step-ahead state-space System Identification (SI). \new{SIMBa leverages}\old{Backed up by novel} linear-matrix-inequality-based free parametrizations of Schur matrices to guarantee the stability of the identified model \emph{by design}\old{, SIMBa allows for seamless integration of prior system knowledge. In particular, it can simultaneously enforce desired system properties --- such as sparsity patterns --- \emph{and} stability on the model, solving an open SI problem}. \new{In this paper, backed up by novel free parametrizations of Schur matrices, we extend the toolbox to show how SIMBa can incorporate}  \new{
known sparsity patterns or true values of the state-space matrices to identify without jeopardizing stability.} 

We extensively investigate SIMBa's behavior when identifying diverse systems with various properties from both simulated and real-world data. Overall, we find it consistently outperforms traditional stable subspace identification methods, and sometimes significantly, \old{even while}\new{especially when} enforcing desired model properties. These results hint at the potential of SIMBa to pave the way for generic structured nonlinear SI. The toolbox is open-sourced on \href{https://github.com/Cemempamoi/simba}{https://github.com/Cemempamoi/simba}.
\end{abstract}

    \section{Introduction}

\IEEEPARstart{N}{eural} Networks (NNs) recently gained a lot of attention, achieving impressive performance on a wide variety of tasks~\cite{goldberg2022neural,rawat2017deep}. Consequently, they have been used for nonlinear System Identification (SI) tasks, where they have also attained state-of-the-art performance~\cite{ljung2020deep, forgione2021continuous, brunton2022data}. While NNs can model complex nonlinearities, they might perform sub-optimally in cases where the structure of the system is known, such as for \textit{linear} systems, where traditional SI methods~\cite{ljung1998system} perform well~\cite{gedon2021deep}.


On the other hand, as demonstrated in~\cite{ecc}, one can leverage the Automatic Differentiation (AD) techniques at the core of NN training to improve upon traditional stable linear SI methods. 
Relying on the efficient and open-source \texttt{PyTorch} Python library~\cite{NEURIPS2019_9015} and a free parametrization of Schur matrices, SIMBa~\cite{ecc} indeed showed significant performance gains compared to traditional stable SI implementations in MATLAB~\cite{ljung1995system} and Python~\cite{armenise2018open}.

In this work, we extend the open-source SIMBa (\textbf{S}ystem \textbf{I}dentification \textbf{M}ethods leveraging \textbf{Ba}ckpropagation) toolbox originally presented in~\cite{ecc} beyond standard stability guarantees, allowing for more general \textit{prior knowledge integration}. Specifically, we show how SIMBa can incorporate structures in the state-space matrices to identify, for example, to enforce desired sparsity patterns. This information might indeed be known to the user \textit{a priori}: one may have insights on which states are measured, which inputs impact which states, or which states exchange information, i.e., the topology of a networked system, among others.

    \subsection{Traditional linear system identification}

Traditionally, input-output state-space models are identified through Subspace Identification Methods (SIMs)~\cite{qin2006overview}. Remarkably, N4SID~\cite{van1994n4sid}, MOESP~\cite{verhaegen1992subspace} and CVA~\cite{larimore1990canonical} have later been unified under a single theory in~\cite{van1995unifying}, proving they rely on similar concepts.
In addition to these classical SIMs, \textit{parsimonious} SIMs (PARSIMs) were introduced more recently to guarantee the causality of the identified models. 
While PARSIM-S~\cite{qin2005novel} and \text{PARSIM-P}~\cite{qin2003parallel} assume no correlation between the input and the output noise, PARSIM-K~\cite{pannocchia2010predictor} was later developed for closed-loop SI.

When stability is desired, \textit{post-hoc} corrections can be applied to the state-space matrices identified by SIMs to obtain a stable model~\cite{maciejowski1995guaranteed, chui1996realization}\new{, and this option is included in most existing implementations~\cite{sjovall2006constrained}, including the well-known MATLAB SI toolbox~\cite{ljung1995system}}. However, this is impossible for PARSIMs and might cause severe performance drops on traditional SIMs~\cite{armenise2018open}. 

Alternatively, one can rewrite the Least Squares (LS) optimization at the heart of SIMs as a \new{regularized~\cite{van2001identification} or} constrained~\cite{boots2007constraint, lacy2003subspace} optimization problem that ensures stability. Similarly, one may leverage parametrization of Schur matrices~\cite{gillis2020note, jongeneel2022efficient} to 
guarantee the stability of the identified system through projected Gradient Descent (GD), for example~\cite{mamakoukas2023learning}.

    \subsection{Prior knowledge integration}

In addition to maintaining the stability of the system, it can be beneficial, and sometimes necessary, to convey expert knowledge or desired properties to the identified model in practice. This led to the development of SIMs specifically tailored for distributed systems with different topologies, for example, where the state-space matrices are known to have specific sparsity patterns~\cite{haber2014subspace,massioni2008subspace,
yu2016subspace}. 
\newre{When dealing with real-world systems with known topologies, such as buildings~\cite{zakwan2022physically}, power networks~\cite{deka2023learning}, and metabolic and regulatory biological networks~\cite{mangan2016inferring}, for example, enforcing the known sparsity on the state-space matrices can indeed help improve the accuracy of the identified models. Incorporating prior topological knowledge into the dynamics also eliminates the possibility of identifying spurious phenomena from data, such as fictitious state dependencies that do not reflect the physical reality.  For instance, while modeling the thermal dynamics of a building, it has been observed that failing to enforce sparsity patterns in the state matrices leads to models that exhibit energy exchange between the rooms that do not share walls~\cite{zakwan2022physically}.} 
\new{SIMs have also been extended to guarantee positive realness, enforce specific model structures (e.g., output-error), or incorporate prior knowledge on the steady-state gain of the system, for example~\cite{van2013closed}.}
\old{More generally, an expert might have prior knowledge about the structure of the system 
stemming from known physical properties, for example. To ensure the identified model follows the desired dynamics,} 
\new{Similarly, to ensure models of physical systems follow known dynamics stemming from the underlying physical laws,}  one typically writes down the corresponding state and output equations manually and then identifies the unknown parameters from data. Such \textit{grey-box} modeling approaches have been successfully applied to building~\cite{li2021grey}, chemical process~\cite{zapf2021gray}, or robotic system~\cite{obadina2022grey} modeling, among others. 

These considerations were recently unified in the COSMOS framework, which allows one to identify structured linear systems from input-output data~\cite{yu2019constrained}. It is based on rewriting the LS estimate in SIMs as a rank-constrained optimization problem to ensure the identified matrices belong to the desired parametrized sets. 
However, COSMOS minimizes the one-step-ahead prediction error and does not guarantee the stability of the identified model. 

We note here that knowledge integration also encompasses the careful initialization of all the parameters, which has already been shown to improve performance in the context of nonlinear SI~\cite{schoukens2021improved}. While this is not needed for classical subspace methods, which rely on deterministic LS solutions~\cite{qin2006overview}, it can have a significant impact on the convergence rate and quality of the solutions of gradient-based algorithms~\cite{dauphin2019metainit}. This was already observed for SIMBa in~\cite{ecc}.


    \subsection{Contribution}

None of the aforementioned methods allowing prior knowledge integration considered the stability of the resulting model. Additionally, they usually rely on one-step-ahead fitting criteria\new{, which can lead to poor long-term prediction accuracy~\cite{farina2011simulation}}. \old{To the best of the authors' knowledge, simultaneously imposing sparsity or other desired properties \textit{and} stability constraints with multi-step-ahead identification is still an open problem that we propose to solve in this paper.} 
\old{To achieve this, we extend SIMBa to let one incorporate desired system properties without jeopardizing the stability of the identified model.}\new{To exemplify how to maintain state-of-the-art performance without losing stability guarantees when enforcing 
desired system properties, this paper extends SIMBa to allow for incorporation of predefined sparsity patterns or known values state-space matrices. }

Leveraging novel free parametrizations of Schur matrices and well-established ML tools for multi-step prediction error minimization, we show how SIMBa can significantly outperform traditional \new{stable} SI methods \new{found in the MATLAB SI toolbox~\cite{ljung1995system}}. To this end, we conduct extensive numerical experiments, identifying different systems with and without state measurements, from simulated as well as real-world data, and while enforcing various system properties. 

SIMBa incurs large computational burdens, requiring from several minutes to over an hour for training, in contrast to the few seconds needed for conventional methods. However, it consistently --- and sometimes significantly --- outperforms traditional approaches on a wide variety of problems in terms of accuracy, even while enforcing \old{desired properties}
\new{prior knowledge on the state-space matrices}. SIMBa could hence be very beneficial in applications where performance is critical or system properties must be respected.

Together, these investigations hint at the efficacy of ML tools to improve upon traditional methods and pave the way for a generic nonlinear SI framework. Indeed, SIMBa could integrate nonlinearities, similarly to~\cite{zakwan2022physically, di2023towards}, or be fused in Koopman-based approaches like~\cite{loyakoopman, schulze2022data, schulze2022identification, choi2023data} to enforce stability.

\textit{Organization:} After a few preliminaries in Section~\ref{sec:preliminaries}, Section~\ref{sec:theory} presents the main theoretical results on the free parametrization of Schur matrices. Section~\ref{sec:simba} then introduces the SIMBa toolbox in detail and Section~\ref{sec:results} provides extensive numerical validations and analyses. Finally, Section~\ref{sec:discussion} discusses some limitations and potential extensions of SIMBa before Section \ref{sec:conclusion} concludes the paper.

    \section{Preliminaries}
    \label{sec:preliminaries}

This work is concerned with the identification of discrete-time linear time-invariant state-space models of the form
\begin{subequations}\label{eq:sys}
\begin{align}
    x_{k+1} &= Ax_k + Bu_k \label{eq:sys1}\\
    y_k &= Cx_k + Du_k \;, \label{eq:sys2}
\end{align}
\end{subequations}
where $x\in\mathbb{R}^n, u\in\mathbb{R}^m,y\in\mathbb{R}^p$ are the states, inputs, and outputs, respectively. 
The objective is to identify $A\in\mathbb{R}^{n \times n}$, $B\in\mathbb{R}^{n \times m}$, $C\in\mathbb{R}^{p \times n}$, and $D\in\mathbb{R}^{p \times m}$ from data. 

To that end, throughout this work, we assume access to a data set $\mathcal{D}^{i/o} = \{\left(u(0), y(0)\right), ..., \left(u(l_s), y(l_s)\right)\}_{s=1}^N$ of $N$ input-output measurement trajectories $s$ of length $l_s$. Note that, in some cases, one might have direct access to state measurements, in which case~\eqref{eq:sys2} is omitted and only $A$ and $B$ need to be identified from a data set of input-state measurements $\mathcal{D}^{i/s}= \{\left(u(0), x(0)\right), ..., \left(u(l_s), x(l_s)\right)\}_{s=1}^N$. In our experiments, we split the data into a training, a validation, and a test set of trajectories $\mathcal{D}_{\textit{train}}$, $\mathcal{D}_{\textit{val}}$, and $\mathcal{D}_{\textit{test}}$, respectively, as often done in ML pipelines~\cite{lones2021avoid}. 

When we want to enforce the asymptotic stability of~\eqref{eq:sys}, we need to ensure $A$ is Schur, i.e., all its eigenvalues  $\lambda_i(A)$ satisfy $|\lambda_i(A)| < 1, \forall i=1,...,n $~\cite{de1999new}. 

Finally, given a matrix $M\in\mathbb{R}^{q\times s}$, the \textit{binary mask} $\text{sparse}(M):=\mathcal{M}\in\{0,1\}^{q\times s}$ represents its \textit{sparsity pattern}. Such a sparse matrix can be written as $M := \mathcal{M} \odot \bar{M}$, with $\bar{M}$ a matrix of appropriate dimensions and $\odot$ the Hadamard product between two matrices.

\new{\textit{Terminology}: Throughout the manuscript, we will say a function is \textit{differentiable} whenever it is the result of a \textit{chain of subdifferentiable operations}, which then allows us to leverage \texttt{PyTorch}'s backpropagation algorithm to \textit{differentiate through it}, i.e., compute its gradient.}

\textit{Notation:} Let $\mathbb{I}_q$ and $\mathds{1}_{q\times q}$ be the identity and all-one matrix of dimension $q$, respectively. Given a matrix $H\in\mathbb{R}^{2q \times 2q}$, we define its block components $H_{11}, H_{12}, H_{21}, H_{22}\in\mathbb{R}^{q \times q}$ as
\begin{equation}
    \begin{bmatrix}
        H_{11} & H_{12} \\ H_{21} & H_{22}
    \end{bmatrix}
    := H\;.
\end{equation}
For a symmetric matrix $F\in\mathbb{R}^{q\times q}$, $F\succ\ 0$ means it is positive definite. For a generic matrix $J\in\mathbb{R}^{q\times q}$, $\lambda_{\textit{min}}(J)$ and $\lambda_{\textit{max}}(J)$ refer to its minimum and maximum eigenvalue, respectively, and
\begin{equation*}
    |\lambda(J)|_{\textit{max}} := \max_{i=1,...,n}|\lambda_i(J)|
\end{equation*}
to its maximum absolute eigenvalue. Finally, $\sigma:= \mathbb{R}\rightarrow ]0,1[$ corresponds to the sigmoid function, i.e., 
$\sigma(r)= (1+e^{-r})^{-1}$, $\sigma^{-1}$ to its inverse, and $||\cdot||_p$ to the $p$-norm of a vector.

    \section{Free Parametrizations of Schur matrices}
    \label{sec:theory}

To efficiently leverage \texttt{PyTorch}'s automatic differentiation framework --- which cannot deal with constrained optimization --- for stable SI tasks, we require $A$ to be Schur \textit{by design}.
This will then allow us to run unconstrained Gradient Descent (GD) in the search space 
without jeopardizing the stability of the identified system. Inspired by~\cite{revay2023recurrent}, we leverage Linear Matrix Inequalities (LMIs) to design matrices that simultaneously guarantee stability and capture various system properties.\footnote{Except for Proposition~\ref{prop:naive}, which relies on different arguments.}

    \subsection{Dense Schur matrices}

For completeness, let us first state a slightly modified version of the free parametrization proposed and proved in~\cite{ecc}, which characterizes Schur matrices with an arbitrary structure.
\begin{proposition}\label{prop:generic}    For any $W\in\mathbb{R}^{2n \times 2n}$, $V\in\mathbb{R}^{n \times n}$, $0< \gamma \leq 1$, and $\epsilon>0$, let 
    \begin{equation}
    S:=W^\top W + \epsilon\mathbb{I}_{2n} \;. \label{eq:S generic}
    \end{equation}
    Then
    \begin{equation} 
        A = S_{12} \left[\frac{1}{2}\left(\frac{S_{11}}{\gamma^2} + S_{22}\right) + V - V^\top\right]^{-1} \label{eq:A generic}
    \end{equation}
    is Schur with 
    $|\lambda_i(A)|<\gamma, \forall i=1,...,n$.
\end{proposition}
Note that $\gamma$ is a user-defined parameter bounding the eigenvalues of $A$ in a circle of the corresponding radius centered at the origin, 
potentially enforcing desired system properties on the learned matrix. This parametrization has been shown to capture \textit{all} Schur matrices $A$ when $\epsilon=\epsilon(A)$ is adapted during training~\cite[Remark 1]{ecc}. This can, however, lead to numerical instabilities in practice if $\epsilon$ takes excessively small values and we hence fix it to a small arbitrary constant in our implementations. Critically, it does not hinder the representation power of Proposition~\ref{prop:generic}, which can still capture all Schur matrices, as detailed in the following corollary.

\begin{corollary}\label{cor:generic}
    For any given Schur matrix $A$ and $\epsilon>0$, there exists $W\in\mathbb{R}^{2n \times 2n}$, $V\in\mathbb{R}^{n \times n}$ satisfying~\eqref{eq:A generic} for $S$ as in~\eqref{eq:S generic}.
\end{corollary}
\begin{proof}
    See Appendix~\ref{app:corgen}.
\end{proof}

\begin{remark}
\new{Recurrent Equilibrium Networks (RENs) employ similar LMI-based arguments for stable nonlinear SI~\cite{revay2023recurrent}. Discarding the nonlinear part of RENs, one could recover a parametrization of stable linear systems resembling SIMBa when leveraging~\eqref{eq:A generic} for stable linear SI. However, RENs do not allow for the incorporation of system properties beyond stability, contrary to SIMBa.}
\end{remark}

    \subsection{Discretized continuous-time systems}
    \label{sec:continuous}

Many linearized real-world systems are continuous-time, i.e., of the form %
\begin{subequations}\label{eq:sys cont}
\begin{align}
    \dot{x} &= \bar{A}x + \bar{B}u \label{eq:sys cont1}\\
    y &= Cx + Du\;. \label{eq:sys cont2}
\end{align}
\end{subequations}
After a forward Euler discretization, \eqref{eq:sys cont} becomes 
\begin{align*}
    x_{k+1} &= \left(\mathbb{I}_n + \delta \bar{A} \right)x_k + \delta \bar{B}u_k \\
    y_k &= Cx_k + Du_k\;,
\end{align*}
where $\delta > 0$ is the discretization step.\footnote{While different discretization schemes exist, we focus on the forward Euler one herein as it allows us to derive another meaningful parametrization of Schur matrices.} In particular, the matrix $A$ we want to identify from discrete-time data samples while ensuring stability now takes the form 
\begin{equation}
    A := \mathbb{I}_n + \delta \bar{A} \;. \label{eq:A}
\end{equation}
In other words, if the data in $\mathcal{D}^{i/o}$ or $\mathcal{D}^{i/s}$ has been collected from a continuous-time system, then $A$ will be \textit{close to identity}. Note that a similar behavior would also be expected from slow-changing systems. The following proposition offers a parametrization of $A$ that takes this structure into account. 

\begin{proposition}\label{prop:continuous}
    For any $W\in\mathbb{R}^{2n \times 2n}$, $V\in\mathbb{R}^{n \times n}$, and $\epsilon>0$, let 
    \begin{equation}
    S := W^\top W + \epsilon\mathbb{I}_{2n}. \label{eq:S continuous}
    \end{equation}
    Then
    \begin{equation}
        A = \mathbb{I}_n - 2\left(S_{11} + V - V^\top\right)^{-1}S_{12}S_{22}^{-1}S_{21} \label{eq:A continuous}
    \end{equation}
    is a Schur matrix.
\end{proposition}
\begin{proof}
    See Appendix~\ref{app:continuous}.
\end{proof}
Contrary to Proposition~\ref{prop:generic} and as evident from~\eqref{eq:A continuous}, the matrix $A$ will be steered towards the identity matrix here, as desired. If the given data stems from a continuous-time or slow-changing discrete-time system, this might ease SIMBa's learning procedure, which we leverage in Section~\ref{sec:res is}. Importantly, Proposition~\ref{prop:continuous} does not sacrifice any representation power, as demonstrated in the following corollary. 

\begin{corollary}\label{cor:continuous}
    For any given Schur matrix $A$ and $\epsilon>0$, there exists $W\in\mathbb{R}^{2n \times 2n}$, $V\in\mathbb{R}^{n \times n}$  satisfying~\eqref{eq:A continuous} for $S$ as in~\eqref{eq:S continuous}.
\end{corollary}
\begin{proof}
    See Appendix~\ref{app:corcont}.
\end{proof}

\begin{remark}
    The discretization step $\delta$ 
    does not appear in~\eqref{eq:A continuous}. While it might seem counter-intuitive at first glance, this is not an issue in practice: changing the discretization step would indeed modify the data collection for $\mathcal{D}^{i/o}$ or $\mathcal{D}^{i/s}$, thereby naturally changing the solution found by SIMBa through GD and hence the form of $A$. In other words, $A$ does implicitly depend on $\delta$, as expected. 
\end{remark}

    \subsection{Sparse Schur matrices}

Let us now assume the sparsity pattern $\mathcal{M}$ of $A$ is given. This might arise in cases where the system to identify is a networked system with a known topology, for example.  
The following proposition shows one possibility to parametrize such sparse matrices without losing stability guarantees.
\begin{proposition}\label{prop:sparse}
    For a given sparsity pattern $\mathcal{M}\in\{0,1\}^{n \times n}$, any $W\in\mathbb{R}^{2n \times 2n}$, $V\in\mathbb{R}^{n \times n}$, and $\epsilon>0$, let 
    \begin{equation}
    S := W^\top W + \epsilon\mathbb{I}_{2n} \label{eq:S sparse}
    \end{equation}
    and construct the diagonal matrix $N$ with entries
    \begin{equation}
    N_{ii} :=  \max \left\{\sum_{j \neq i} \mathcal{M}_{ij}, \sum_{j \neq i} \mathcal{M}_{ji} \right\} + \epsilon,\ \forall i=1,...,n \;. \label{eq:N sparse}
    \end{equation}
    Then, the matrix
    \begin{equation}
        A = \mathcal{M} \odot \left( S_{12} \left[ N \odot \left( \frac{1}{2} \left(S_{11} + S_{22}\right) + V - V^\top \right) \right]^{-1} \right) \label{eq:A sparse}
    \end{equation}
    is Schur and presents the desired sparsity pattern $\mathcal{M}$.
\end{proposition}
\begin{proof}
    See Appendix~\ref{app:sparse}.
\end{proof}

For a given sparsity pattern $\mathcal{M}$ and small positive constant $\epsilon$, one can thus define $N$ and use the free parametrization~\eqref{eq:A sparse} to compute Schur matrices presenting the desired sparsity pattern from some $V$ and $W$. 


\begin{remark}\label{rem:sparse}
    Contrary to Propositions~\ref{prop:generic} and~\ref{prop:continuous}, Proposition~\ref{prop:sparse} is conservative; it \emph{cannot} capture all sparse Schur matrices. This stems from two steps in Appendix~\ref{app:sparse}. First, we have to restrict our search to systems admitting diagonal Lyapunov functions to leverage the associative property of Hadamard products with diagonal matrices. 
    Second, satisfying~\eqref{eq:levy}--\eqref{eq:sparse LMI2} is only a sufficient condition for~\eqref{eq:app sparse} to hold.
\end{remark}

\old{\begin{remark}\label{rem:cons}
    Setting $\mathcal{M}:=\mathds{1}_{n\times n}$ would provide another parametrization of dense Schur matrices, potentially replacing Proposition~\ref{prop:generic}. However, this is not advised in practice since~\eqref{eq:A sparse} is more conservative than~\eqref{eq:A generic} (see Remark~\ref{rem:sparse}). 
\end{remark}}

    \subsection{An alternative general parametrization}

To showcase the power of \texttt{PyTorch}, which can differentiate through the computation of eigenvalues, the following proposition offers an alternative free parametrization of any Schur matrix. Contrary to the other parametrizations, it relies on scaling arguments instead of LMIs.

\begin{proposition}\label{prop:naive}
    For a given sparsity pattern $\mathcal{M}\in\{0,1\}^{n \times n}$, $0< \gamma \leq 1$, and any matrix $V\in\mathbb{R}^{n \times n}$ and constant $\eta\in\mathbb{R}$, leveraging the sigmoid function $\sigma$, the matrix
    \begin{equation}
        A = \frac{\sigma(\eta)\gamma}{|\lambda(\mathcal{M} \odot V)|_{\textit{max}}} \left(\mathcal{M} \odot V\right) \label{eq:A naive}
    \end{equation}
    is Schur with 
    $|\lambda_i(A)|<\gamma, \forall i=1,...,n$, and presents the desired sparsity pattern $\mathcal{M}$.
\end{proposition}
\begin{proof}
    See Appendix~\ref{app:naive}.
\end{proof}

As can be seen, Propositions~\ref{prop:sparse} and~\ref{prop:naive} can be used interchangeably; they both provide a free parametrization of sparse and stable matrices. Contrary to its conservative counterpart, however, Proposition~\ref{prop:naive} can capture \textit{all} Schur matrices -- including sparse ones --- as shown in the following corollary.

\begin{corollary}\label{cor:naive}
    Any Schur matrix $A$ satisfies~\eqref{eq:A naive} for some $\mathcal{M}\in\{0,1\}^{n \times n}$, $0<\gamma\leq 1$, $V\in\mathbb{R}^{n \times n}$, and $\eta\in\mathbb{R}$.
\end{corollary}
\begin{proof}
    One can set $\mathcal{M}:=\text{sparse}(A)$,  
    $V := A$, $\gamma$ such that $|\lambda(A)|_{\textit{max}} < \gamma \leq 1$, which exists since the matrix $A$ is Schur, and $\eta := \sigma^{-1}\left( |\lambda(A)|_{\textit{max}}/\gamma\right)$, which is well-defined since $|\lambda(A)|_{\textit{max}}/\gamma<1$ by definition of $\gamma$.
\end{proof}

Interestingly, defining $\mathcal{M}:=\mathds{1}_{n\times n}$ in Proposition~\ref{prop:naive}, we recover a free parametrization of generic matrices, providing an alternative to Proposition~\ref{prop:generic}. \old{However, according to Corollary~\ref{cor:naive}, this would not come at the cost of expressiveness, contrary to 
Proposition~\ref{prop:sparse}. }Similarly, parametrizing $V$ as in~\eqref{eq:A} and using $\mathcal{M}:=\mathds{1}_{n\times n}$, we recover a parametrization of matrices close to identity interchangeable with Proposition~\ref{prop:continuous}. Overall, Proposition~\ref{prop:naive} hence allows us to characterize \textit{any} type of Schur matrix discussed throughout this Section.

\begin{remark}
    Proposition~\ref{prop:naive} is conceptually related to~\cite{kolter2019learning}, where a Lyapunov function is learned simultaneously to nominal system dynamics. 
    At each step, the dynamics are then projected onto the Lyapunov function to guarantee asymptotic stability. 
    Similarly, \eqref{eq:A naive} can be seen as a projection onto some (unknown) Lyapunov function. However, the latter is implicitly defined through the scaling of $A$ instead of being learned, hence alleviating the associated computational burden.
\end{remark}

\begin{remark}
    Leveraging techniques similar to~\cite{martinelli2023unconstrained}, Propositions~\ref{prop:generic} and~\ref{prop:sparse} can be adapted for continuous-time systems of the form~\eqref{eq:sys cont}. On the other hand, the scaling approach deployed in Proposition~\ref{prop:naive} to control the magnitude of the eigenvalues of $A$ cannot be straightforwardly adapted to the continuous-time setting, where the real part of each eigenvalue has to be negative to ensure stability.
\end{remark}

    \subsection{Using free parametrizations}
    \label{sec:learning}

Propositions~\ref{prop:generic}--\ref{prop:naive} imply one can choose \textit{any} $\epsilon>0$, $0<\gamma\leq 1$, $\mathcal{M}\in\{0,1\}^{n \times n}$, $V\in\mathbb{R}^{n\times n}$, $W\in\mathbb{R}^{2n\times 2n}$, and $\eta\in\mathbb{R}$ --- depending on the setting --- and construct a stable matrix $A$ as in~\eqref{eq:A generic}, \eqref{eq:A continuous}, \eqref{eq:A sparse}, or~\eqref{eq:A naive}. In practice, $\epsilon$ should be set to a small constant,\footnote{We use $\epsilon=\texttt{1e-6}$ in our experiments.} and $\gamma$ and $\mathcal{M}$ are problem-specific and user-defined since they stem from prior knowledge about the system. Therefore, all \textit{constrained} parameters are defined by the user \textit{a priori}.

SIMBa then searches for $V$, $W$, and $\eta$ optimizing some performance criterion as detailed in Section~\ref{sec:simba}. Since these parameters are \textit{not} constrained, we can use unconstrained GD for this task. Propositions~\ref{prop:generic}--\ref{prop:naive} hence allow us to leverage the full power of \texttt{PyTorch} 
to fit the data without jeopardizing stability, 
constructing a Schur matrix $A$ from the \textit{free parameters} $V$, $W$, and $\eta$ at every iteration.

    \section{The SIMBa toolbox}
    \label{sec:simba}

\begin{figure}
    \begin{center}
    \includegraphics[width=0.9\columnwidth]{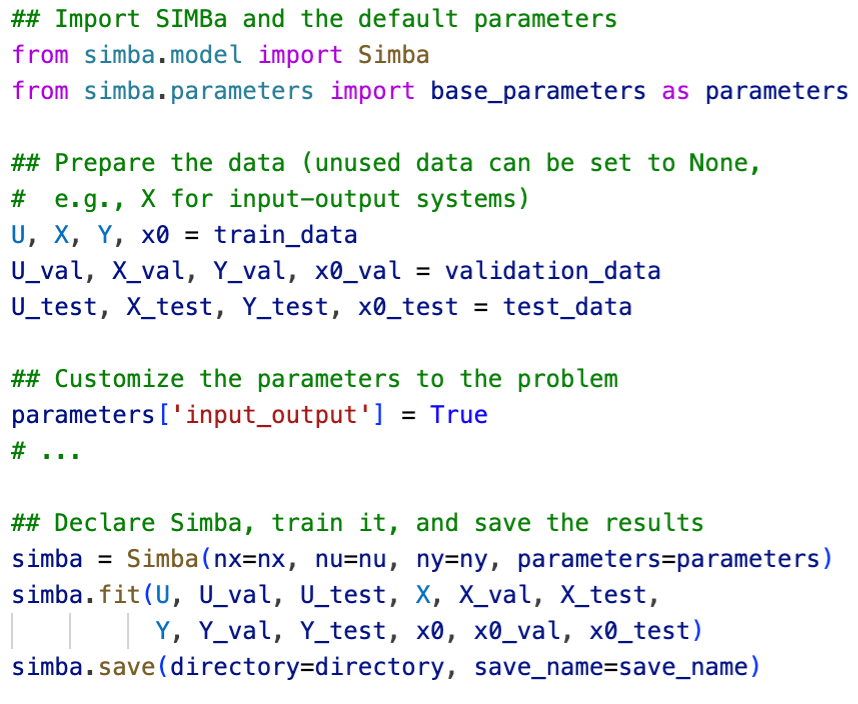}
    \caption{Main steps of running SIMBa in Python.}
    \label{fig:python}
    \end{center}
\end{figure}

This Section describes the main \texttt{parameters} of SIMBa, restating the ones discussed in~\cite{ecc} for completeness, introducing new ones pertaining to prior knowledge integration, and discussing critical hyperparameters in more detail. SIMBa is implemented in Python to leverage the efficient AD framework of \texttt{PyTorch}~\cite{NEURIPS2019_9015} and open-sourced on \href{https://github.com/Cemempamoi/simba}{https://github.com/Cemempamoi/simba}. For given data sets $\mathcal{D}_{\textit{train}}$, $\mathcal{D}_{\textit{val}}$, and $\mathcal{D}_{\textit{test}}$, SIMBa can be initialized, fit, and saved in a few lines of codes, as exemplified in Fig.~\ref{fig:python}. We also provide a MATLAB interface inspired by the traditional MATLAB SI toolbox~\cite{ljung1995system}\old{, as shown in Fig.~\ref{fig:matlab}}.

    \subsection{Optimization framework}
    \label{sec:io systems}

Given input-output data $\mathcal{D}^{i/o}$, SIMBa iteratively runs gradient descent on batches of trajectories $Z\in\mathcal{D}_{\textit{train}}$ randomly sampled from the training data set --- thus seamlessly handling training data sets consisting of several trajectories --- 
to solve the following optimization problem:
\begin{align}
    \min_{A,B,C,D,x_{0}^{(s)}} &\quad \frac{1}{|Z|}\sum_{s\in Z}\left[\frac{1}{l_s}\sum_{k=0}^{l_s} m_k^{(s)}\mathcal{L}_{\textit{train}} \left( y^{(s)}(k), y^{(s)}_k \right)\right] \label{eq:obj io}\\
    \text{s.t.} &\quad  y^{(s)}_k = Cx^{(s)}_k + Du^{(s)}(k) \label{eq:y}\\
                &\quad  x^{(s)}_{k+1} = Ax^{(s)}_k + Bu^{(s)}(k) \;. \label{eq:x}
\end{align}
In words, SIMBa minimizes the multi-step-ahead prediction error, using the training loss $\mathcal{L}_\textit{train}$ as performance criterion. 
In this paper, we rely on the Mean Squared Error (MSE), i.e., $\mathcal{L}_\textit{train}(y,\hat{y})=||y-\hat{y}||_2^2$. However, SIMBa's flexibility --- backed by \texttt{PyTorch}'s ability to handle any differentiable function --- allows one to design custom (differentiable) loss functions and pass them through the \texttt{train\_loss} parameter. For some applications, it might for example be interesting to optimize the Mean Absolute Error (MAE) or the Mean Absolute Percentage Error (MAPE), which are respectively more robust against outliers or different output magnitudes.

In many cases, identifying the matrix $D$ is not required, which is achieved in SIMBa by setting \texttt{id\_D=False}, removing the second term of~\eqref{eq:y}. Similarly, if $x^{(s)}(0)$ is known, \texttt{learn\_x0} can be toggled to \texttt{False} and SIMBa will fix $x^{(s)}_0:=x^{(s)}(0)$ instead of optimizing it.
The number of sequences $|Z|$ used for each gradient update can be controlled through the \texttt{batch\_size} parameter. 

Finally, $m_k^{(s)}\in\{0,1\}$ in~\eqref{eq:obj io}, with
\begin{align*}
    m_k^{(s)} = \begin{cases}
        0, &\text{with probability } p \text{ or if } y^{(s)}(k) \text{ is \texttt{NaN}}, \\
        1, &\text{otherwise}.
    \end{cases}
\end{align*}
In words, these binary variables let SIMBa discard missing values from the objective function, allowing it to seamlessly work with incomplete data sets. 
Additionally, the user can define a \texttt{dropout} $=p$ parameter to randomly remove data points from the objective with probability $p$, providing empirical robustness to the training procedure. Indeed, it can be seen as a means to avoid overfitting the training data.

    \subsection{Input-state identification}
    \label{sec:is systems}
    
When state measurements are available in $\mathcal{D}^{i/s}$, one can set \texttt{input\_output=False}, dropping 
\eqref{eq:y}, and modifying the objective to 
\begin{align}
    \min_{A,B} &\quad \frac{1}{|Z|}\sum_{s\in Z}\left[\frac{1}{l_s}\sum_{k=0}^{l_s} m_k^{(s)}\mathcal{L}_{\textit{train}} \left( x^{(s)}(k), x^{(s)}_k \right)\right],\label{eq:obj is}
\end{align}
\new{where $Z$ contains randomly sampled trajectories from $\mathcal{D}^{i/s}_{\textit{train}}$.}
Furthermore, for autonomous systems, the \texttt{autonomous} flag can be toggled, in which case the minimization on $B$ is also discarded, as well as the corresponding second term on the right-hand-side of~\eqref{eq:x}. Similarly to Section~\ref{sec:io systems}, $m_k^{(s)}$ are binary variables that can be forced to zero either to discard missing values or as a means of regularization.

Since the state $x$ is known, one can break given training trajectories into \textit{segments} of length \texttt{horizon}. 
The \texttt{stride} defines how many steps should be taken between the \textit{starts} of two segments. Note that if \texttt{stride} is smaller than \texttt{horizon}, then segments of data will overlap, i.e., data points will appear several times in consecutive segments. If the user is interested in the model performance over a specific horizon length, this can be specified with \texttt{horizon\_val}, and the number of segments can also be controlled with \texttt{stride\_val}. Note that setting \texttt{horizon} or \texttt{horizon\_val} to \texttt{None} keeps entire trajectories.

    \subsection{Training procedure}
    \label{sec:training}

SIMBa iteratively 
runs one step of gradient descent on~\eqref{eq:obj io} or~\eqref{eq:obj is} for \texttt{max\_epochs} epochs. Here, we define an \textit{epoch} as one pass through the training data, i.e., every trajectory has been drawn in a batch $Z$. 
After each epoch, the validation data is used to assess the current model performance by computing
\begin{equation*}
    \frac{1}{|\mathcal{D}_\textit{val}|}\sum_{s\in \mathcal{D}_\textit{val}}\left[\frac{1}{l_s}\sum_{k=0}^{l_s} \mathcal{L}_{\textit{val}} \left( y^{(s)}(k), y^{(s)}_k \right)\right] \;,
\end{equation*}
replacing outputs by states when measurements are available. This evaluation metric is used to avoid SIMBa overfitting the training data: the best parameters, stored in memory, are only overwritten if the current parameters improve the performance on the validation set, and not the training one~\cite{lones2021avoid}. At the end of the training, to get a better estimate of the true performance of the identified model, we evaluate it on the unseen test data:
\begin{equation*}
    \frac{1}{|\mathcal{D}_\textit{test}|}\sum_{s\in \mathcal{D}_\textit{test}}\left[\frac{1}{l_s}\sum_{k=0}^{l_s} \mathcal{L}_{\textit{val}} \left( y^{(s)}(k), y^{(s)}_k \right)\right]\;,
\end{equation*}
similarly substituting outputs by states when measurements are available. Throughout this work, we also rely on the MSE for validation and testing, setting $\mathcal{L}_\textit{val}=\mathcal{L}_\textit{train}$, but this can be modified through the \texttt{val\_loss} parameter. This gives the users the freedom to evaluate SIMBa on a different metric than the training one. It can be tailored to specific applications: SIMBa would then return the model performing best with respect to the chosen evaluation criterion, irrespective of the training procedure. 

Note that~\eqref{eq:obj io} or~\eqref{eq:obj is} can be highly nonconvex, in which case gradient descent cannot be expected to find the global optimum and will most likely settle in a local one instead. SIMBa is thus sensitive to its initialization and some hyperparameters and might converge to very different solutions depending on these choices.

     \subsection{Initialization}

To start SIMBa in a relevant part of the search space, the user can toggle \texttt{init\_from\_matlab\_or\_ls} parameter to \texttt{True}. This prompts SIMBa to a run traditional SI method, i.e., either
\begin{itemize}
    \item the MATLAB SI toolbox~\cite{ljung1995system} or the Python \texttt{SIPPY} package~\cite{armenise2018open} for input-output systems,\footnote{Since several traditional SI methods are available, SIMBa uses the one achieving the best performance on the validation set.} or
    \item a traditional LS optimization in the input-state case,
\end{itemize} 
before training. Depending on the setting, the chosen \textit{initialization method} returns matrices $A^*$, and potentially $B^*$, $C^*$, and $D^*$. These are then used as initial choices of state-space matrices in SIMBa, so that it starts learning from the best solution found by traditional SI methods.

However, to ensure the often-desired stability of $A$, SIMBa relies on the free parametrizations in Propositions~\ref{prop:generic}--\ref{prop:naive}, in which case it is not possible to directly initialize $A$ to $A^*$. We thus again resort to \texttt{PyTorch} to approximate it by solving the following optimization problem with unconstrained GD:
\begin{align}
    \min_{W,V,\eta} &\quad \mathcal{L}_{\textit{init}}\left(A, A^*\right) \label{eq:obj A}\\
    \text{s.t.} &\quad  A\text{ as in~\eqref{eq:A generic}, \eqref{eq:A continuous}, \eqref{eq:A sparse}, or~\eqref{eq:A naive}}\;, \label{eq:cstnt A}
\end{align}
with $\mathcal{L}_\textit{init}$ the desired loss function. We use the MSE throughout our numerical experiments but custom functions can be passed through the \texttt{init\_loss} parameter.

Similarly to what was mentioned in Section~\ref{sec:training} concerning SIMBa's training, this procedure is not guaranteed to find the global optimal solution, i.e., to converge to $A^*$. Consequently, even initialized instances of SIMBa might perform very differently from the initialization method, sometimes incurring large performance drops. Nonetheless, throughout our experiments, it usually worked well, finding an $A$ close to $A^*$. \old{, as long as no prior system knowledge is available.}

\old{Indeed, since desired system properties cannot be enforced on traditional SI methods, the initialization method will always return dense and generic state-space matrices. 
In that case, SIMBa would try to match matrices following the required patterns to these generic matrices, which might lead to poor solutions. It is hence not advised to use \texttt{init\_from\_matlab\_or\_ls} when prior system knowledge is available; one should rather directly enforce this knowledge on SIMBa and let it learn from scratch.}

\begin{remark}
    Note that $A$ might be initialized exactly in the specific case when Proposition~\ref{prop:naive} is leveraged for stability and $\gamma>|\lambda(A^*)|_\textit{max}$, as detailed in the proof of Corollary~\ref{cor:naive}. On the other hand, although Corollaries~\ref{cor:generic}--\ref{cor:continuous} demonstrate that any Schur matrix $A^*$ can be parametrized as~\eqref{eq:A generic} or~\eqref{eq:A continuous}, respectively, they only show the \emph{existence} of such a parametrization and cannot be used in practice to construct it.
\end{remark}

    \subsection{Prior knowledge integration}

If certain sparsity patterns are desired for $A$, $B$, $C$, or $D$, they can be passed through \texttt{mask\_\{X\}}, replacing \texttt{\{X\}} with the name of the corresponding matrices. If the mask of $B$ is given as $\mathcal{M}_B$, for example, \eqref{eq:x} is modified to
\begin{align}
    x^{(s)}_{k+1} = Ax^{(s)}_k + (\mathcal{M}_B \odot B)u^{(s)}(k) \label{eq:mask}
\end{align}
to force the desired entries of $B$ to zero while letting SIMBa learn the others.

Similarly, \texttt{\{X\}\_init} is used to initialize a given matrix to a specific value, and \texttt{learn\_\{X\}=False} drops the corresponding matrix from the optimization, fixing it at its initial value. To control the magnitude of the eigenvalues of $A$ in the free parametrizations of Propositions~\ref{prop:generic} and~\ref{prop:naive}, one can set \texttt{max\_eigenvalue} $=\gamma$. 

Finally, \texttt{stable\_A=True} enforces the stability of $A$: setting \texttt{naive\_A=True} leverages Proposition~\ref{prop:naive} while toggling \texttt{LMI\_A} uses Propositions~\ref{prop:generic}--\ref{prop:sparse}. Specifically, if \texttt{delta} is not \texttt{None} but takes the value $\delta$, hinting we are expecting $A$ to be close to the identity matrix, then Proposition~\ref{prop:continuous} is used instead of Proposition~\ref{prop:generic}. Similarly, if \texttt{mask\_A} is not \texttt{None}, hence requiring a sparse system, Proposition~\ref{prop:sparse} is leveraged. 

When \texttt{stable\_A=True}, the minimization over $A$ in~\eqref{eq:obj io} or~\eqref{eq:obj is} is replaced by a minimization over $W$, $V$, and/or $\eta$ --- depending on which of the four Propositions is used --- and constraint~\eqref{eq:cstnt A} is added to the corresponding optimization problem. 


    \subsection{Tuning of critical hyperparameters}


As for NNs, which heavily rely on the same backpropagation backbone, the \texttt{learning\_rate} is an important parameter: too large values lead to unstable training while too small ones slow the convergence speed. In general, throughout our empirical analyses, the default value of \texttt{1e-3} showed very robust performance, and we couple it with a high number of epochs to ensure SIMBa can converge close to a local minimizer. However, this is problem-dependent, and the rate might be increased to accelerate learning if no instability is observed. Similarly, \texttt{init\_learning\_rate} controls the learning rate of the initialization in~\eqref{eq:obj A}--\eqref{eq:cstnt A} when required. It also defaults to \texttt{1e-3} due to its robust performance when coupled with a high number of \texttt{init\_epochs}.

To promote stable learning, we implemented a gradient clipping operation, pointwise saturating the gradients of all the parameters to avoid taking overly aggressive update steps during GD. These can be controlled through \texttt{grad\_clip} and \texttt{init\_grad\_clip} during the training and the initialization phase, respectively. The default values of \texttt{$100$}, respectively \texttt{$0.1$}, were empirically tuned to achieve good performance. Although it might come at the cost of a slower convergence, our numerical investigations showed gradient clipping can significantly improve the quality of the solution found by SIMBa.

    \section{Numerical experiments}
    \label{sec:results}

As described in Section~\ref{sec:simba}, SIMBa can identify models from input-output and input-state measurements while seamlessly enforcing desired system properties such as stability or \old{sparsity patterns}\new{prior knowledge on the state-space matrices}. This Section provides numerical examples showcasing its ability to outperform traditional \new{stable} SI methods 
in a wide variety of case studies. It exemplifies how SIMBa can leverage Propositions~\ref{prop:generic}--\ref{prop:naive} to guarantee the stability of the identified model while achieving state-of-the-art fitting performance. Interestingly, our investigations hint that \new{enforcing known sparsity patterns or true values of state-space matrices}\old{integrating prior knowledge, while being easy,} does not impact the quality of the solution found by SIMBa in general. On the contrary, it seems that domain knowledge injection can be helpful to improve performance.

Sections~\ref{sec:res random}--\ref{sec:res naive} first analyze the behavior of SIMBa on \old{different }simulated input-output data sets, complementing the results\old{ obtained} in~\cite{ecc}. Throughout these experiments, we fixed $x_0=0$, which allows us to compare SIMBa with \texttt{SIPPY}'s implementations of SIMs and PARSIMs in Section~\ref{sec:res random}~\cite{armenise2018open}.\footnote{In these cases where $x_0=0$, the results found by MATLAB's SI toolbox~\cite{ljung1995system} turned out to be either comparable or slightly worse than \texttt{SIPPY}'s solutions. They are thus not reported herein. Note that MATLAB's SI toolbox would outperform \texttt{SIPPY} when $x_0\neq 0$, as discussed in~\cite{ecc}.} \new{
Sections~\ref{sec:res structure}--\ref{sec:res naive} subsequently compare SIMBa's performance to the one of the MATLAB SI toolbox~\cite{ljung1995system} when known sparsity patterns or true values of state-space matrices are enforced.} 

We then investigate the performance of SIMBa on an input-state identification task from real-world measurements in Section~\ref{sec:res is}. We show how SIMBa surpassed the standard LS method and the state-of-the-art SOC approach for stable SI from~\cite{mamakoukas2020memory}\new{, leveraging Proposition~\ref{prop:continuous} to identify a stable discretized continuous-time system}. Finally, Section~\ref{sec:time} details the computational burden associated with SIMBa.\footnote{The code and data used for these experiments can be found on \href{https://gitlab.nccr-automation.ch/loris.dinatale/simba}{https://gitlab.nccr-automation.ch/loris.dinatale/simba}.}

    \subsection{Comparing Propositions~\ref{prop:generic} and~\ref{prop:naive} using random systems}
    \label{sec:res random}

First, to complement the results in~\cite{ecc} showing the superiority of SIMBa on randomly generated systems when leveraging Proposition~\ref{prop:generic} to guarantee stability, we use the same simulation setting here to compare the performance when Proposition~\ref{prop:naive} is used instead. Note that both free parametrizations can capture \textit{all} stable matrices and thus find the true solution (see Corollaries~\ref{cor:generic} and~\ref{cor:naive}).

To that end, we generated $50$ stable discrete-time systems as in~\cite{ecc}, with one training, one validation, and one testing trajectory of $300$ steps, $n=5$, $m=3$, and $\new{p}\old{r}=3$. The inputs were generated as a Generalised Binary Noise (GBN) input signal with a switching probability of $0.1$ for each dimension~\cite{armenise2018open} and white output noise $\mathcal{N}(0, 0.25)$ was added on the training trajectory. For each system, we defined two instances of SIMBa, one with \texttt{LMI\_A=True} (\textit{SIMBa-1}) and another with \texttt{naive\_A=True} (\textit{SIMBa-4}), leveraging the corresponding free parametrizations.

\begin{figure}
    \begin{center}
    \includegraphics[width=\columnwidth]{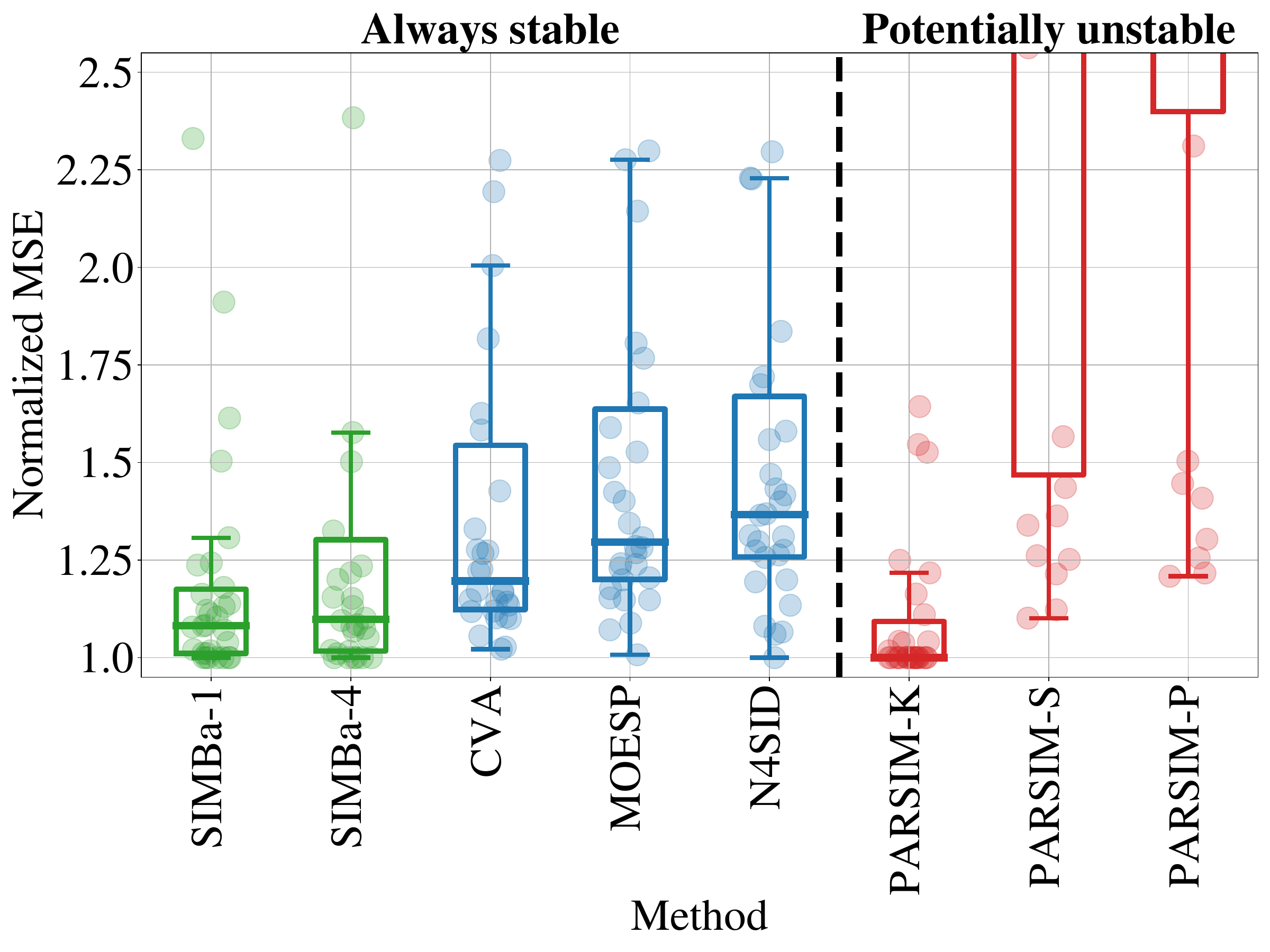}
    \caption{Performance of input-output state-space identification methods on $30$ randomly generated systems, where the MSEs have been normalized by the best-obtained error for each system. The performance of SIMBa (ours) --- either relying on the parametrization proposed in Proposition~\ref{prop:generic} or~\ref{prop:naive} --- is plotted in green, other stable SI methods in blue, and unstable ones in red.}
    \label{fig:random naive}
    \end{center}
\end{figure}

First, the performance of each SI method on the testing trajectory from $30$ systems is plotted in Fig.~\ref{fig:random naive}, \new{where the clouds of points correspond to the accuracy achieved on each system. For comparison with SIMBa (green), the errors of state-of-the-art stable SIMs and PARSIMs, implemented in \texttt{SIPPY}~\cite{armenise2018open}, are reported in blue and red, respectively}\old{where green indicates SIMBa, blue other stable SI methods, and red PARSIM methods, which cannot guarantee stability}. For each system, the MSE of every method is normalized with respect to the best-attained performance. 

We observe similar performance between both parametrizations, with a slight edge on Proposition~\ref{prop:generic}. It achieved a median MSE $8.2\%$ worse than the best method on the different systems, compared to the $9.8\%$ of \textit{SIMBa-4}, as reported in Table~\ref{tab:random naive} for clarity. \new{\textit{SIMBa-4} additionally shows a slightly less robust performance than \textit{SIMBa-1}, with a wider interquartile range. This hints that while the free parametrization in Proposition~\ref{prop:naive} does capture all stable matrices, it might underperform empirically compared to the one from Proposition~\ref{prop:generic}.}

Note that PARSIM-K attained impressive performance in this setting --- it is the only traditional SI method able to compete with SIMBa in terms of accuracy, confirming the trend observed in~\cite{ecc}. 
However, \old{it}\new{PARSIMs} cannot guarantee the stability of the identified model.\footnote{\new{Over the $50$ systems identified in Section~\ref{sec:res random}, PARSIM-K, PARSIM-S, and PARSIM-P returned zero, eight, and twelve unstable models, respectively.}} On the other hand, \old{other stable SI methods reached}\new{stable SIMs obtained} a performance drop of more than $20\%$ compared to the best method half of the time\old{ (see Table~\ref{tab:random naive})}.

\begin{table}[!t]
    \caption{Performance drop of each method compared to the best one, reported from Fig.~\ref{fig:random naive}.}
    \label{tab:random naive}
    \centering
    \begin{tabular}{l|c|c|c}
    \hline
    \textbf{Method} & \textbf{$0.25$-quantile} & \textbf{Median} & \textbf{$0.75$-quantile} \\
    \hline
    SIMBa-1		&	\textbf{1.01}	&	\textbf{1.08}	&	\textbf{1.18}	\\
    SIMBa-4		&	1.02	&	1.10	&	1.30	\\
    CVA		&	1.12	&	1.20	&	1.54	\\
    MOESP		&	1.20	&	1.30	&	1.64	\\
    N4SID		&	1.26	&	1.37	&	1.67	\\ \hline
    PARSIM-K	&	\textbf{1.00}	&	\textbf{1.00}	&	\textbf{1.09}	\\
    PARSIM-S	&	1.47	&	4.75	&	30.91	\\
    PARSIM-P	&	2.40	&	7.00	&	223.76	\\
    \hline
    \end{tabular}
\end{table}

\old{Note that \textit{SIMBa-4} additionally shows a slightly less robust performance than \textit{SIMBa-1}, with a wider interquartile range. This hints that while the free parametrization in Proposition~\ref{prop:naive} does capture all stable matrices, it might be numerically less stable than the LMI-based one from Proposition~\ref{prop:generic}.}

\begin{figure}
    \begin{center}
    \includegraphics[width=\columnwidth]{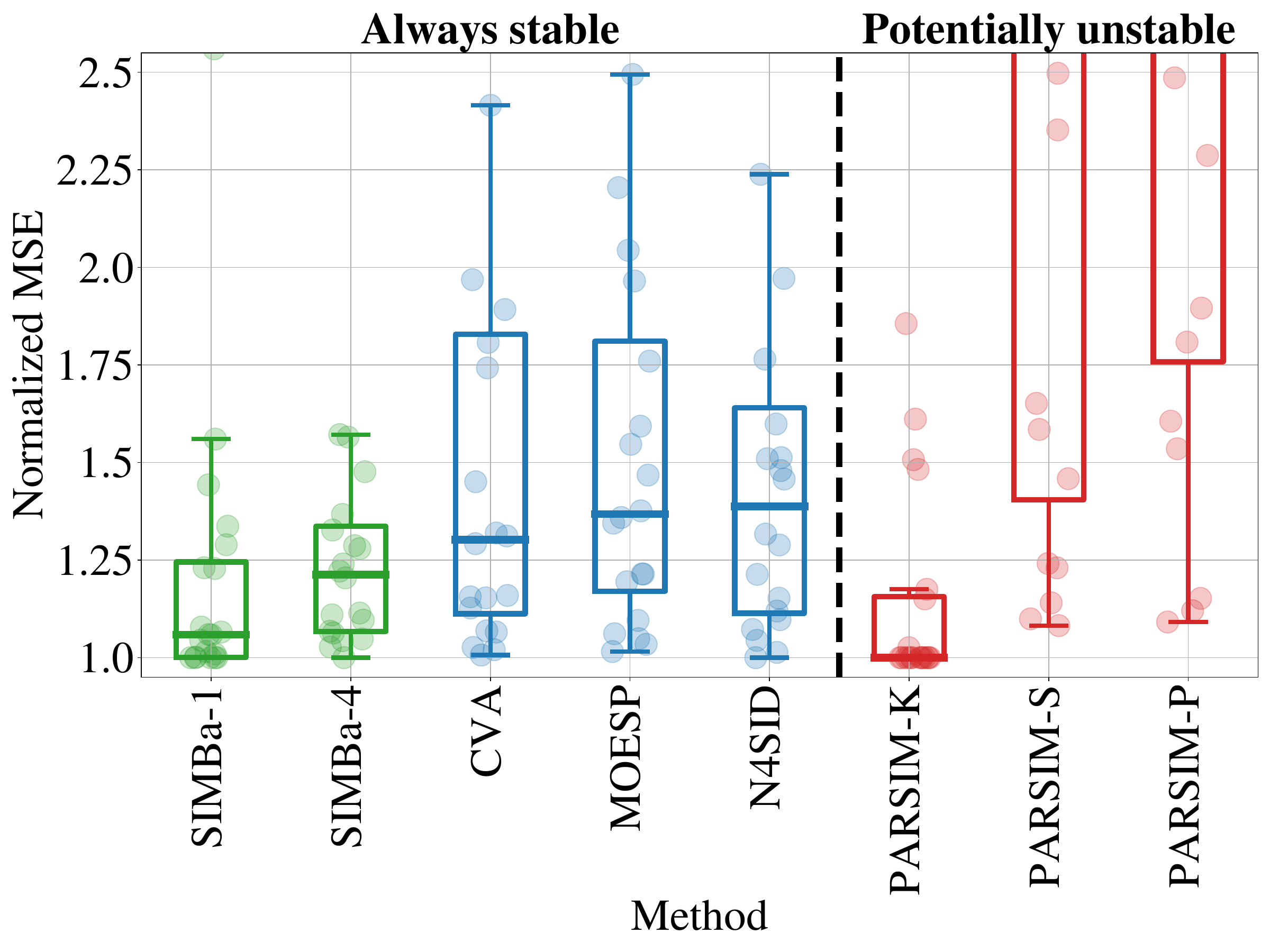}
    \caption{Performance of input-output state-space identification methods on $20$ randomly generated systems, where the data has been standardized and the MSEs have been normalized by the best-obtained error for each system. The performance of SIMBa (ours) --- either relying on the parametrization proposed in Proposition~\ref{prop:generic} or~\ref{prop:naive} --- is plotted in green, other stable SI methods in blue, and potentially unstable ones in red.}
    \label{fig:random naive norm}
    \end{center}
\end{figure}

Finally, for completeness, we used the other $20$ generated systems to assess the impact of data standardization on the final performance. Before the SI procedure, each dimension of the dataset was processed to have zero mean and unit standard deviation, removing the effect of different dimensions having different magnitudes, as is often done in practice. 
As pictured in Fig.~\ref{fig:random naive norm}, however, little impact can be seen, with the different methods reaching similar performance to what was observed in Fig.~\ref{fig:random naive}. On the contrary, there seems to be a slightly wider gap between \textit{SIMBa-1} and the other stable SI approaches in blue. Similarly to the previous case, \text{\textit{SIMBa-4}} again slightly underperformed compared to its counterpart leveraging Proposition~\ref{prop:generic}, providing additional indications that the parametrization in Proposition~\ref{prop:naive} might be numerically more challenging. \new{However, further case studies would be required to confirm this hypothesis in other settings.}

Since little performance difference can be observed between Fig.~\ref{fig:random naive} and Fig.~\ref{fig:random naive norm}, data standardization does not seem to impact SIMBa's performance significantly in general. Interestingly, this means even the gradient-based SIMBa can be run to fit data with different orders of magnitudes accurately. We suspect gradient clipping to be an important reason behind this strong performance, but further analyses would be required to understand the behavior of GD in SIMBa fully.

    \subsection{\old{Introducing prior knowledge}\new{Enforcing known system properties}}
    \label{sec:res structure}

As a second case study, \old{we analyzed}\new{this Section analyzes} the effect of incorporating various levels of prior knowledge \new{--- i.e., enforcing known sparsity patterns or true values of one or several of the state-space matrices ---} into SIMBa \new{without jeopardizing stability}. 
To that end, we used the same simulation settings as in Section~\ref{sec:res random} to create $10$ stable systems but with $n=7$, $m=6$, $p=5$, and trajectories of length $500$. Before generating the data, however, we randomly set $60\%$ of the entries of $A$, $B$, $C$, and $D$ to zero.\footnote{We made sure that $A$ remained stable after this sparsification procedure.}\old{ We used these sparse systems to assess the impact of prior knowledge, i.e., either knowing the sparsity pattern or the true values of one or several of the state-space matrices.} We let SIMBa run for \numprint{25000} epochs. To ensure stability, we set \texttt{LMI\_A=True} to leverage Proposition~\ref{prop:sparse} or~\ref{prop:generic} when \texttt{mask\_A} is known or not, respectively, \new{or \texttt{naive\_A=True} to use Proposition~\ref{prop:naive}.\footnote{\new{The used Proposition is encoded in the name of each SIMBa's instance.}} When the latter parametrization was leveraged to identify sparse Schur matrices, we additionally ran several instances of SIMBa to assess the impact of randomness --- the random seed and initialization of the trainable parameters --- on its performance, and we report here the median and minimum error achieved on each generated system. This shows what can be expected on average but also the best attainable performance with \textit{SIMBa-4}. For comparison, the same tasks were solved with the \texttt{ssest} function from the MATLAB SI toolbox~\cite{ljung1995system}}. 

\begin{figure}
    \begin{center}
    \includegraphics[width=\columnwidth]{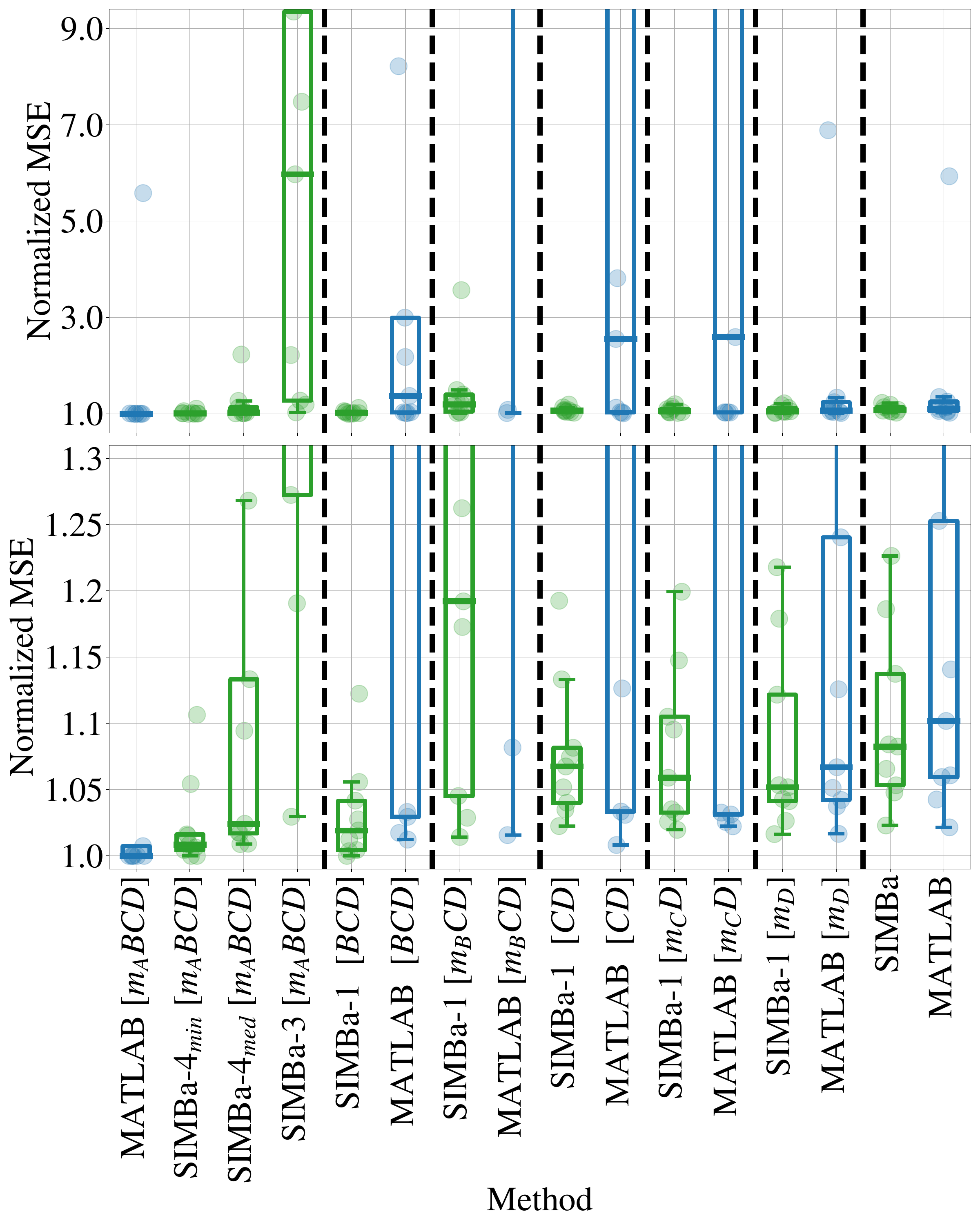}
    \caption{\new{Normalized MSE of each method on test input-output data from $10$ randomly generated systems with sparse matrices $A$, $B$, $C$, and $D$. The letters in square brackets encode which matrices $X$ or sparsity pattern $m_X$, respectively, are assumed to be known and fixed. Both plots show the same data with a different zoom to appreciate the difference between SIMBa (ours) --- either relying on Proposition~\ref{prop:generic}, \ref{prop:sparse} or~\ref{prop:naive} --- in green and the \texttt{ssest} function in the MATLAB SI toolbox in blue. 
    \textit{SIMBa-4} was run with 
    eight random seeds on each system, and we report both the median and minimum error.}}
    \label{fig:structure}
    \end{center}
\end{figure}

The resulting normalized errors on the testing data are presented in Fig.~\ref{fig:structure}, where the plots have been generated as in Section~\ref{sec:res random}, and the bottom figure is a zoomed-in version for better visualization of the differences between the various instances of SIMBa. The \old{knowledge}\new{known system properties} incorporated in each SIMBa \new{or MATLAB} instance is encoded \new{in square brackets} in their name, where ``$X$'' or ``$m_X$'' indicates that the true matrix $X$ or its true sparsity pattern was given through \texttt{\{X\}\_init} or \texttt{mask\_\{X\}}, respectively.\footnote{When the true matrix $X$ is given to SIMBa, \texttt{learn\_\{X\}} is set to \texttt{False}, so that it is not modified during learning.} For example, \new{\textit{[$m_BCD$]}} represents instances with knowledge of $C$, $D$, and the sparsity pattern of $B$. In practice, this could correspond to a system where $D\equiv 0$ and we know which states are measured (i.e., $C$ is known) and which inputs act on which states but not their exact impact (i.e., the sparsity pattern of $B$ is known). This is encoded in SIMBa by setting \texttt{learn\_C=learn\_D=False}, passing the known matrices $C$ and $D$ as \texttt{C\_init} and \texttt{D\_init}, respectively, and defining \texttt{mask\_B} to be the true known sparsity pattern of $B$.

\old{Since prior knowledge cannot be enforced in any traditional method, they are only directly comparable with the vanilla system-agnostic \textit{SIMBa}. As expected, similar conclusions as in Section~\ref{sec:res random} and in~\cite{ecc} can be drawn from the top plot of Fig.~\ref{fig:structure}, with a clear edge of \textit{SIMBa} over other stable SI methods and only PARSIM-K achieving comparable performance. Indeed, traditional stable SI methods lead to relatively poor performance, never coming closer than $20\%$ off the best performance and incurring costs over $1.5$--$1.75\times$ higher than the best one half of the time. Furthermore, as shown in the bottom plot, PARSIM-K achieved a median gap to the best performance of around $10\%$, which is worse than most informed SIMBa instances despite not guaranteeing stability.}

\old{Interestingly, apart from \textit{SIMBa\_DCBmA} --- when simultaneously enforcing stability and sparsity of the matrix $A$ with Proposition~\ref{prop:sparse} ---, all the informed instances of SIMBa consistently achieve strong performance improvements over traditional stable methods. Most notably, they often outperformed the system-agnostic \textit{SIMBa}, hinting at the efficacy of prior knowledge inclusion in SIMBa.}
\new{In general, increasing 
the amount of system knowledge (from right to left in Fig.~\ref{fig:structure}) is positively reflected in SIMBa's performance when leveraging Propositions~\ref{prop:generic} or~\ref{prop:naive} in this case study. This hints at the efficacy of system properties incorporation 
in SIMBa, except for \textit{SIMBa-1 [$m_BCD$]}.\footnote{\new{Interestingly, MATLAB also struggled to converge to meaningful solutions in this setting, hinting that fitting a Schur $A$ and selected entries of $B$ was not trivial in this experiment.}}} On the one hand, this makes intuitive sense since we pass \textit{true} information to SIMBa, restricting the search space. On the other hand, enforcing fixed matrices or sparsity patterns reduces the expressiveness of the model to fit the training data well --- and GD might get stuck in a poor local minimum. Indeed, the set of all possible state-space matrices, over which \textit{SIMBa} optimizes,\footnote{More specifically, it optimizes over the space of stable matrices $A$ and generic matrices $B$, $C$, and $D$.} contains the search spaces 
the other instances are optimizing over. \textit{SIMBa} might hence find linear systems attaining a better test accuracy 
without respecting the desired system properties. Although \textit{SIMBa-3 [$m_ABCD$]}\old{and, to some extent, \textit{SIMBa-1 [$m_BCD$]}} seems to have been impacted and stuck in \old{global}\new{local} minima, the other informed versions of SIMBa all 
\new{achieved errors within $25\%$ of the best one in almost all cases}. 

\new{In contrast, MATLAB only achieved reasonable accuracy in three of the seven experiments, when either no or little prior knowledge was enforced (on the right of Fig.~\ref{fig:structure}), or when only the values of $A$ with known sparsity pattern needed to be learned from data (on the left of the plot). These results provide evidence that enforcing desired system properties, even if our assumptions are correct, might deteriorate MATLAB's performance significantly. On the contrary, SIMBa converged to accurate solutions throughout our experiments, typically significantly outperforming standard SI methods. 
}

\new{The only setting where MATLAB outperformed SIMBa was when only selected entries of $A$ had to be learned, assuming all other matrices to be known ($m_ABCD$). Nevertheless, as can be seen in the top of Fig.~\ref{fig:structure}, MATLAB 
did not converge to a meaningful solution on one system, only obtaining a testing accuracy more than five times lower than the best one. In contrast, \textit{SIMBa-4$_{min}$ [$m_ABCD$]} showed a more consistent performance across the different systems, never achieving an error more than $11\%$ off the best one.}


Altogether, these investigations hint that SIMBa indeed allows one to impose known or desired system properties without sacrificing significant model performance in general\new{, contrary to MATLAB}. There is however one critical exception: Proposition~\ref{prop:sparse} seems to impose overly conservative conditions on sparse \new{Schur} matrices, in line with Remark~\ref{rem:sparse}, and often led to poor accuracy in this case study~(\textit{SIMBa-3 [$m_ABCD$]}). 

    \subsection{\old{Identifying sparse and stable systems}\new{Identifying sparse Schur matrices}}
    \label{sec:res naive}

\begin{figure}
    \begin{center}
    \includegraphics[width=\columnwidth]{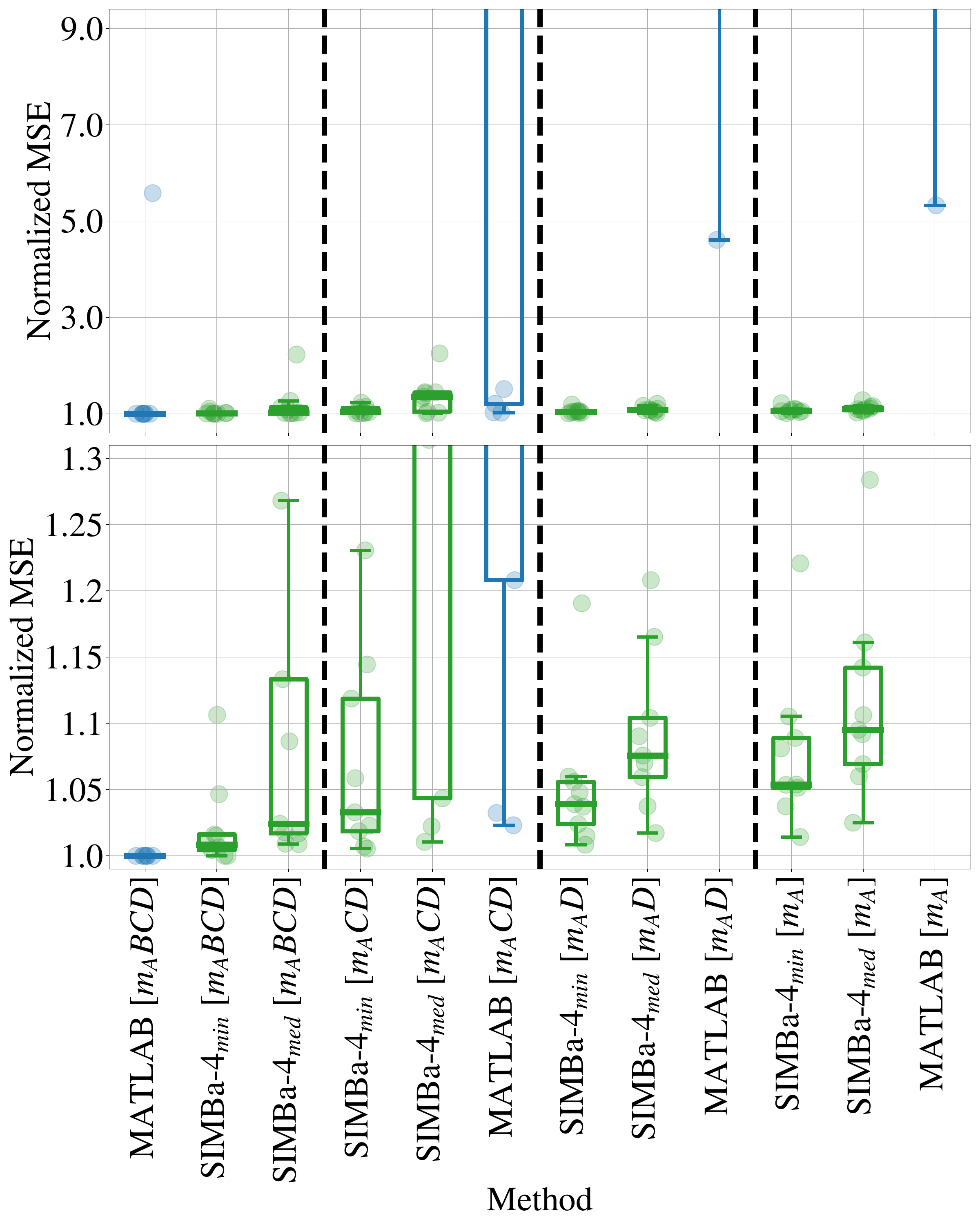}
    \caption{\new{Performance of each method, normalized by the best one, on test input-output data from $10$ randomly generated systems with sparse matrices $A$, $B$, $C$, and $D$. The performance of SIMBa (ours) --- relying on Proposition~\ref{prop:naive} --- is plotted in green, the one of the \texttt{ssest} function in the MATLAB SI toolbox in blue, and the bottom plot is a zoomed-in version of the top one for better visualization}. Note that SIMBa was run with $10$ different random seeds on each system, and we report both the median and minimum error.} 
    \label{fig:mask}
    \end{center}
\end{figure}

\old{To tackle the aforementioned issue of Proposition~\ref{prop:sparse} being too conservative and constraining the expressiveness of SIMBa, we propose to use Proposition~\ref{prop:naive} instead in this Section.} \new{
To complement Section~\ref{sec:res structure} and assess the efficacy of Proposition~\ref{prop:naive} in parametrizing sparse Schur matrices, this Section offers another set of more challenging identification experiments, where 
Schur $A$ matrices with known sparsity patterns have to be identified simultaneously to other state-space matrices. To that end,} we used the same simulation settings \new{and ten systems} as in Section~\ref{sec:res structure}\new{, again 
running several randomly initialized instances of \textit{SIMBa-4} and reporting both the corresponding median and best performance}.\old{ to generate $10$ new sparse systems and compare the effectiveness of both propositions. In other words, we only considered \textit{SIMBa\_DCBmA} here, but using two different free parametrizations of sparse Schur matrices. Furthermore, to ensure randomness did not have a significant impact on the results, we let \textit{SIMBa-4} run with seven different random seeds in each case.} 
All the results on the testing data are plotted in Fig.~\ref{fig:mask} \new{, with the bottom plot being a zoomed-in version to better appreciate the impact of prior knowledge on SIMBa.} 
\old{To better assess the potential of SIMBa, Fig.~\ref{fig:mask min} displays the results when only considering the best-performing seed on each system,
and the corresponding key metrics are reported in Table~\ref{tab:mask min} for clarity. }

\old{As expected from what could be observed in Fig.~\ref{fig:structure}, \text{\textit{SIMBa-3}} performed very poorly in general, even if it was run several times. Interestingly, this analysis hints that the effectiveness of Proposition~\ref{prop:sparse} depends more on system-intrinsic characteristics than on the random seed used. Indeed, although not fully visible in the plots of Fig.~\ref{fig:mask} and~\ref{fig:mask min}, the performance distribution of \textit{SIMBa-3} shows similar statistics whether all the results or only the best seeds are considered. This eliminates the effect of randomness to some extent and supports the hypothesis that the parametrization in~\eqref{eq:A sparse} is too conservative (see Remark~\ref{rem:sparse}).} 

\old{On the other hand, \textit{SIMBa-4} achieved excellent performance in general, as shown in Fig.~\ref{fig:mask}, outperforming every other stable SI method, with PARSIM-K being the only one attaining similar accuracy. In fact, $75\%$ of its instances achieved a performance gap below $30\%$ while all the other stable SI methods in blue had a gap worse than that more than $75\%$ of the time. However, it seems to be more sensitive to randomness, with some instances achieving relatively poor performance, especially compared to the very consistent results obtained by PARSIM-K.} 
\new{As can be seen, apart from the $m_ABCD$ case already discussed in the previous Section, where MATLAB could attain lower errors, SIMBa always 
achieved significantly better performance. 
In fact, MATLAB failed to converge to meaningful solutions in all the other experiments, consistently producing errors that were often orders of magnitude more severe than the best result attained by SIMBa. Notably, while the quality of the solutions found by {MATLAB} significantly deteriorates with the number of parameters to identify simultaneously to the sparse Schur matrices $A$ (from left to right in Fig.~\ref{fig:mask}), SIMBa could keep similarly low errors in all cases. Furthermore, in line with the observations made in the previous Section, we can observe a positive correlation between SIMBa's accuracy and the quantity of pre-existing knowledge about the state-space matrices to identify.} 

\new{On the other hand, SIMBa showed sensitivity to randomness on some systems, with the minimum error achieved being distinctly lower than the median one.} Interestingly, this \new{sensitivity} did not seem to \old{be system-dependent}\new{significantly impact SIMBa's best performance. Indeed,} \textit{SIMBa-4$_{min}$} could achieve low testing error on all the systems\new{, with its accuracy remaining within $22\%$ of the best one in all cases and within $10\%$ three times out of four.}\old{, as plotted in Fig.~\ref{fig:mask min}.}
Although running several instances of SIMBa naturally incurs additional computational cost, \old{Fig.~\ref{fig:mask min}}\new{these results} hint that it can be beneficial to find a better-performing model. \old{If only the best seed is considered, SIMBa might even 
consistently outperform PARSIM-K. In this case study, \textit{SIMBa-4} could simultaneously beat all other stable methods by more than $25\%$ on all systems.}

Collectively, these investigations show that introducing prior knowledge on the sparsity of $A$ does not necessarily come at the price of performance, complementing what was observed in Section~\ref{sec:res structure} for other knowledge integration schemes. \new{Altogether, this shows how SIMBa can conform to system properties desired by the user without significantly suffering in terms of performance.}

    \subsection{Performance on real-world input-state data}
    \label{sec:res is}

To showcase SIMBa's versatility, we now turn to an input-state data set collected from the Franka Emika Panda robotic arm and provided in~\cite{mamakoukas2020memory} \new{for linear input-state identification}. We have access to eight trajectories of length $N=400$, samples at \SI{50}{\hertz}, with $n=17$ and $m=7$. Specifically, the $17$ states of the robots consists of the $x$, $y$, and $z$ coordinates of the end effector and the angle and angular velocity of the seven joints. The seven control inputs correspond to joint velocities. We fixed one validation and one test trajectory, respectively, and used a subset of the remaining six trajectories as training data. Since the robot is a continuous-time system, we leveraged Proposition~\ref{prop:continuous}, setting \texttt{delta=$\frac{1}{50}$} and \texttt{LMI\_A=True}. As we are not dealing with input-output data anymore, subspace identification methods perform poorly, and we hence compare SIMBa to LS and its state-of-the-art stable version, SOC, proposed in~\cite{mamakoukas2020memory}. 
\new{For a fair comparison, since SOC cannot incorporate prior system knowledgse in the identification procedure, we assume no knowledge of the robot's physics being available to SIMBa.}

Here, we also used the ability of SIMBa to work with batched data, breaking the training trajectories into $10$-step long segments, overlapping at each time step, --- i.e., setting \texttt{horizon=10} and \texttt{stride=1} --- to facilitate training for this more complex problem. This gave rise to approximately \numprint{400} to \numprint{2400} training sequences of $10$ steps whether one to six trajectories were used for training. Since we are interested in the final performance of SIMBa over entire trajectories, we set \texttt{horizon\_val=None} to select the best model accordingly. We let instances initialized with the LS solution (\textit{SIMBa\_i}) and randomly initialized ones (\textit{SIMBa}) run for \numprint{20000} and \numprint{40000} epochs, respectively, with a batch size of \numprint{128}. 

\begin{figure}
    \begin{center}
    \includegraphics[width=\columnwidth]{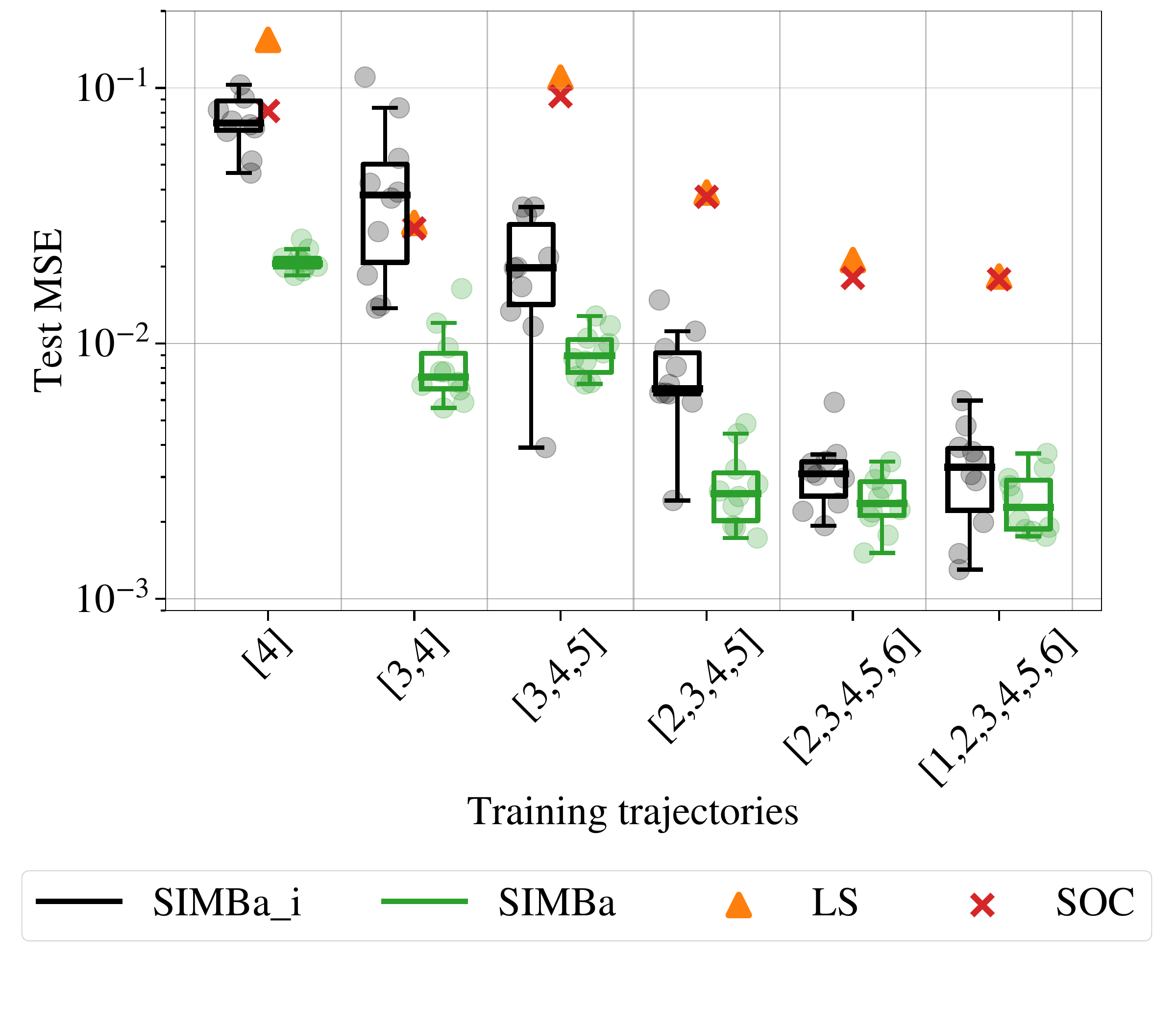}
    \caption{MSE of each method on the test trajectory of the Franka data set after training from different trajectories. Black and green data show the performance of SIMBa over $10$ runs with informed (\textit{SIMBa\_i}) and random initialization (\textit{SIMBa}), respectively. The MSE of LS and SOC are reported in orange triangles and red crosses, respectively.}
    \label{fig:franka}
    \end{center}
\end{figure}

The MSE of each method on the test trajectory is reported in Fig.~\ref{fig:franka} with a logarithmic scale, where the x-axis enumerates which trajectories were used for training. Similarly to the previous Section, SIMBa was run $10$ times in each case to assess the impact of randomness. 
As expected, we see a general tendency of all the methods to find more accurate solutions with more training data. Interestingly, \textit{SIMBa} often performs better than \textit{SIMBa\_i} on this data, hinting that initializing SIMBa with the matrices found through LS might get it stuck in relatively poor local minima.

\begin{figure}
    \begin{center}
    \includegraphics[width=\columnwidth]{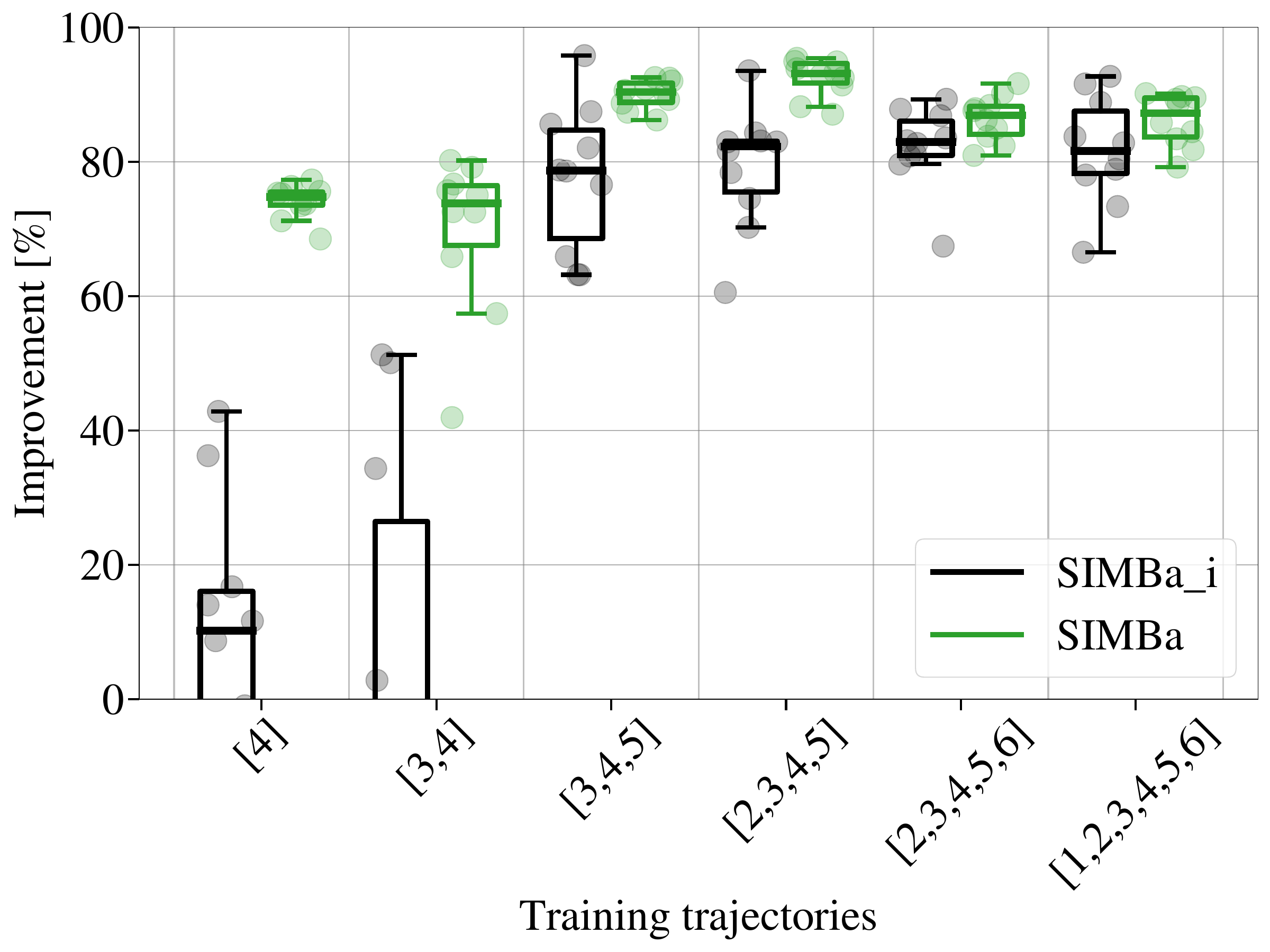}
    \caption{Improvement of SIMBa over SOC on the test data from the Franka robotic arm for different training trajectories, reported from Fig.~\ref{fig:franka}.}
    \label{fig:franka improvement}
    \end{center}
\end{figure}

Overall, SIMBa generally outperformed LS and SOC, and often significantly, especially when more training data was available. The only exceptions came from \textit{SIMBa\_i} when one or two trajectories only were used for training, where we can see performance drops for some instances. On the other hand, \textit{SIMBa} always outperformed LS and SOC, and with an impressive median performance improvement compared to SOC of over $70\%$ and as high as $95\%$, as reported in Fig.~\ref{fig:franka improvement}. Here, the \textit{improvement} is computed as 
\begin{equation*}
    \text{Improvement} = 100\left(1-\frac{\text{MSE}_\textit{SIMBa}}{\text{MSE}_\textit{SOC}}\right) \;.
\end{equation*}

Moreover, looking at the best-achieved performance of \textit{SIMBa} and \textit{SIMBa\_i}, they attained improvements of over $90\%$ compared to SOC as soon as more than three training trajectories were used. When only one or two trajectories were leveraged for training, \textit{SIMBa\_i} achieved $42\%$ or $51\%$ better performance than SOC, respectively, and this improvement increases to $77\%$ or $80\%$ for \textit{SIMBa}.

We suspect the observed performance gaps to be heavily impacted by the ability of SIMBa to minimize the error over \textit{multiple steps}, here ten, compared to myopic classical LS-based methods. This showcases the usefulness of backpropagation-based approaches, which can handle complex fitting criteria instead of the classical one-step-ahead prediction error. 

    \subsection{Training complexity}
    \label{sec:time}

To conclude these numerical investigations, 
this section provides insights into the training time of different versions of SIMBa. All the experiments were run on a Bizon ZX5000 G2 workstation. Note that while a Graphical Processing Unit (GPU) interface is implemented, we did not use it to obtain the results presented in this paper, setting \texttt{device='cpu'}. Indeed, using GPUs for such small-scale problems generally slows the overall training time since the overhead required to move data and models to the GPU at each iteration is higher than the subsequent optimization time gain. 

First, as reported in Table~\ref{tab:time}, each instance of SIMBa ran for slightly less than \SI{1}{\hour} in Section~\ref{sec:res structure}, which is several orders of magnitude slower than the few seconds required to fit traditional SI methods\new{, typically using the MATLAB SI toolbox}. Interestingly, however, enforcing prior knowledge did not significantly impact the run time. Since fewer parameters need to be learned from data, informed versions tended to take less time per epoch. This stems from SIMBa's architecture and unconstrained training procedure in~\eqref{eq:obj io} or~\eqref{eq:obj is}. It allows SIMBa to seamlessly guarantee the desired system properties through modifications of~\eqref{eq:y}--\eqref{eq:x} --- for example, to~\eqref{eq:mask} --- without additional computational burden. 

\begin{table}[!t]
    \caption{\new{Average running time in seconds of each method in Section~\ref{sec:res structure}}.}
    \label{tab:time}
    \centering
    \begin{tabular}{r|c|c}
    \hline
    \textbf{Prior} & \multirow{2}{*}{\textbf{MATLAB}} & \multirow{2}{*}{\textbf{SIMBa}} \\
    \textbf{knowledge} &  & \\
    \hline
    --	&	1.67 &	3729	\\
    $m_D$	&	1.05   &	3713		\\
    $m_CD$	&	0.96	&	3512 \\ 
    $CD$    &   0.82    &	3388		\\
    $m_BCD$	&	0.83	&   3671 \\
    $BCD$	&	0.55	&	3130	\\ 
    $m_ABCD$	&	0.28	&	3100	\\ \hline
    \end{tabular}
\end{table}

Note that these SIMBa instances were run for \numprint{25000} epochs, and doubling that number would hence double their training time to approximately \SI{2}{\hour}. The training procedure would then be comparable to \textit{SIMBa\_L} in~\cite{ecc}, which was reported to take roughly \SI{25}{\minute} for \numprint{50000} epochs, i.e., around five times less. However, the latter was trained over $100$ data points, compared to $500$ in Section~\ref{sec:res structure}, revealing an approximately linear relationship between the horizon length and training complexity.

\begin{figure}
    \begin{center}
    \includegraphics[width=\columnwidth]{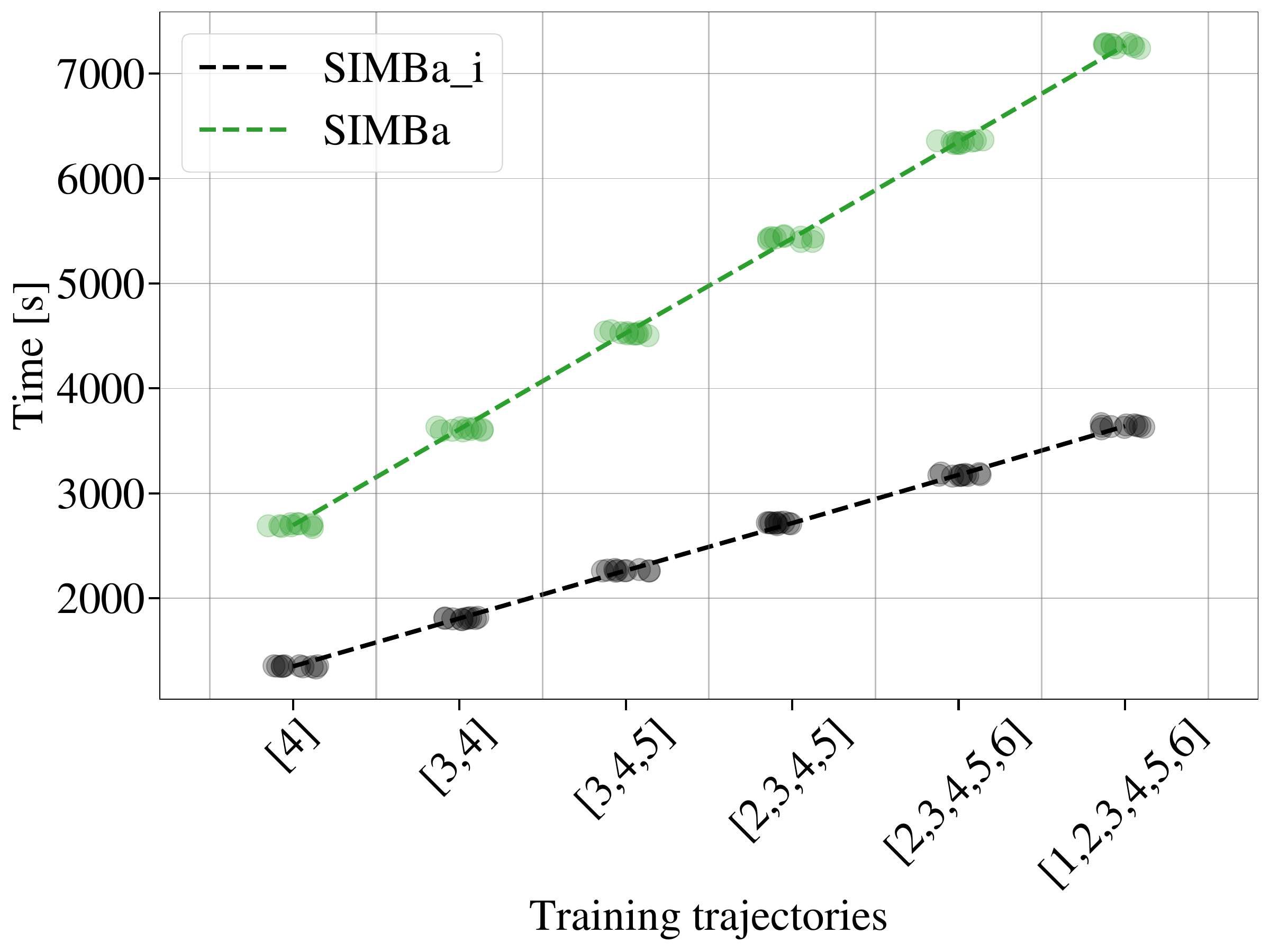}
    \caption{\old{Running time of }\new{Time required by the different instances of} SIMBa (in seconds) \new{to obtain the results} reported in Fig.~\ref{fig:franka}. Note that this does not include approximately \SI{130}{\second} needed to initialize \textit{SIMBa\_i} for \numprint{150000} epochs.}
    \label{fig:times}
    \end{center}
\end{figure}

For completeness, the training times of the SIMBa instances analyzed in Section~\ref{sec:res is} are shown in Fig.~\ref{fig:times}, exposing the expected linear impact of leveraging more and more training data. However, five to six times more data can be used before doubling the training time, leveled by \texttt{PyTorch}'s capability to process several trajectories in parallel. Unsurprisingly, on the other hand, doubling the number of iterations approximately yields twice longer fitting times (comparing \textit{SIMBa\_i} to \textit{SIMBa}).

\begin{figure}
    \begin{center}
    \includegraphics[width=\columnwidth]{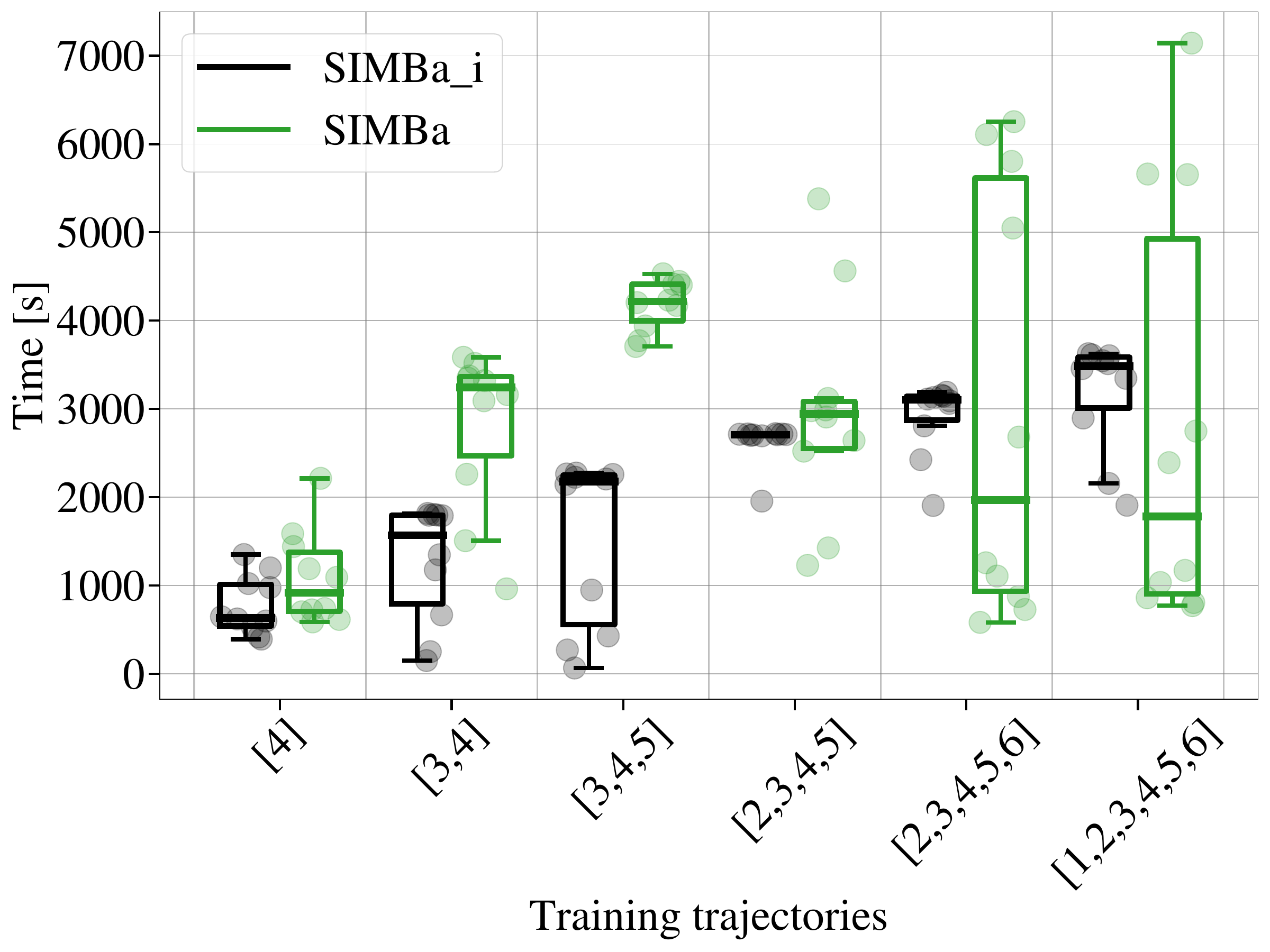}
    \caption{Time required for SIMBa to achieve its best validation error (in seconds), to compare with the corresponding total training times in Fig.~\ref{fig:times}.} 
    \label{fig:real times}
    \end{center}
\end{figure}

As a final remark, we would like to highlight here that all the analyzed instances of SIMBa were usually run for more epochs than required to ensure convergence. In practice, the training could be interrupted once the validation error stagnates or augments, showing SIMBa started to overfit the training data. To illustrate this, Fig.~\ref{fig:real times} reports the time required by SIMBa to achieve its best performance on the validation data set --- these state-space matrices are the ones ultimately employed to assess its performance on the testing data. As can be seen, SIMBa sometimes identifies the best-performing solution in considerably less time than the total allowed training time. Similarly, the learning rate could be increased in practice to converge faster to these solutions; however, a comprehensive analysis of its influence was out of the scope of this \new{study}\old{manuscript}.

    \section{Discussion}
    \label{sec:discussion}

With the investigations in the previous Section showcasing SIMBa's performance on various system identification tasks, let us now briefly summarise how to best use the toolbox to maximize its performance and discuss potentially interesting extensions to it. 

    \subsection{A summary of SIMBa's capabilities}

While SIMBa can achieve significant performance gains over traditional methods in many different settings, the associated computational burden can be significant. Indeed, even simple systems take several minutes to be fit, and this can grow significantly with longer training trajectories or when more epochs are required, as detailed in Section~\ref{sec:time}. Despite its ability to leverage out-of-the-box ML tools and GPUs, SIMBa is hence not well-suited for problems that demand fast solutions. In practice, traditional methods, which can be trained in a matter of seconds, are thus generally a good starting point. On the other hand, for cases where achieving the best performance is critical or when desired system properties need to be preserved, SIMBa can be extremely beneficial, 
as presented in Section~\ref{sec:results}. 

During our investigations, we saw that randomness could play a significant role; running several instances of SIMBa might greatly improve its performance. Although initializing its state matrices with the one found by classical methods usually accelerates convergence, we also observed cases where randomly initialized versions achieve better final performance. 
Both options should hence be considered in practice.

The free parametrizations from Propositions~\ref{prop:generic}--\ref{prop:naive} can be leveraged to guarantee the stability of the identified model. While Proposition~\ref{prop:naive} can characterize any Schur matrix, it seems to be more sensitive to randomness and more numerically challenging than the other parametrizations. In practice, Propositions~\ref{prop:generic} and~\ref{prop:continuous} should thus be preferred, and only the too-conservative Proposition~\ref{prop:sparse} should be discarded in favor of Proposition~\ref{prop:naive}. 

In general, the provided default parameters perform well --- they have indeed demonstrated robust performance across the variety of case studies analyzed in Section~\ref{sec:results}. To maximize the effectiveness of SIMBa, one might consider increasing the number of training epochs, for instance. On the other hand, the learning rate of $0.001$ chosen in this work 
is sometimes slower than required. This default option proved to be robust across various tasks, but it might be possible to accelerate learning by taking larger parameter updates at each step on some problems.

    \subsection{Potential extensions}

Throughout Section~\ref{sec:results}, we did not strain to obtain the best performance on each case study but focused on fair comparisons between different methods under identical conditions. In practical applications, one should combine SIMBa --- or any SI method --- with data processing procedures, such as standardization, detrending, or filtering, to improve performance. In general, an interface with MATLAB's SI toolbox, which comes with many useful helper functions, would be an interesting extension to SIMBa.


Apart from better integration of such existing tools, to facilitate training on nonconvex long-horizon objectives, \textit{curriculum learning} could be adopted. In this framework, the training starts with the minimization of the one-step-ahead prediction error and then gradually increases the prediction horizon toward the desired one~\cite{bengio2015scheduled}. This could help SIMBa in the early stage of training, accelerating the first iterations by simplifying the problem --- similar in spirit to the proposed initialization from the solution of classical SI methods --- before leveraging the full power of automatic backpropagation for long-horizon optimization. \new{Alternatively, one could apply tools from nonlinear system identification to convexify the training problem and facilitate learning~\cite{umenberger2018convex}.}

Interestingly, Proposition~\ref{prop:naive} could also be used to generate affinely parametrized Schur matrices, similar to those examined in~\cite{yu2018identification}, for example. The generality of this free parametrization could thus allow SIMBa to guarantee stability while enforcing desired properties on $A$ beyond specific sparsity patterns. \new{Proposition~\ref{prop:naive} could also be modified to ensure the identification of Schur matrices close to the identity matrix, thereby parametrizing $A$ matrices corresponding to Euler discretizations of continuous-time systems (similarly to Section~\ref{sec:continuous}). This would allow the user to enforce prior knowledge of the continuous-time system, typically stemming from the laws of physics.} \new{Similarly, minor code modifications would allow one to seamlessly enforce other properties on $B$, $C$, and $D$, 
fixing only parts of their entries, for example.}


In parallel with these efforts to enforce additional system properties, including nonlinearities may be crucial for some applications. Indeed, linear models might not be flexible enough to fit more complex systems. Thanks to the AD backbone used by SIMBa, NNs can be seamlessly added on top of the linear model, learning patterns that are not well-captured by the linear part, such as in~\cite{di2023towards}. Otherwise, inspired by the SINDy toolbox~\cite{brunton2016discovering}, if the class of nonlinearities impacting the dynamics are known, one could extend the state description to $f(x)\in\mathbb{R}^{n'}$, for example, including polynomials like $f(x) = [x^\top, (x^2)^\top]^\top$, and then fit a linear model of the form $x_{k+1} = Af(x_k) + Bu(k)$. Finally, a more cumbersome approach would be to discard the linear framework altogether, write custom dynamics, and then leverage automatic backpropagation as proposed herein to find the required parameters, similar to what was proposed in~\cite{zakwan2022physically} for irreversible port-Hamiltonian systems.



To conclude this discussion, we want to point out a potential link to Koopman-based approaches like~\cite{loyakoopman}, where traditional SI methods were used to identify linear models in the corresponding lifted space. Thanks to SIMBa's construction relying on unconstrained GD, the lifting functions could also be learned simultaneously with the lifted linear model, similar in spirit to~\cite{schulze2022identification, schulze2022data, choi2023data}, potentially improving the accuracy of the end-to-end pipeline without jeopardizing stability. 


    \section{Conclusion}
    \label{sec:conclusion}

In this paper, we extended the SIMBa toolbox to allow for prior system knowledge integration into the identified state-space matrices, going beyond standard stability conditions. Leveraging novel free parametrizations of Schur matrices to ensure the stability of the identified model despite enforcing \old{desired properties, such as sparsity,}\new{predefined sparsity patterns or known values of the state-space matrices,} we showed how SIMBa outperforms traditional \new{stable} SI methods, and often by more than $25\%$. \old{Notably, this performance gap grew to $50\%$ when SIMBa had access to prior knowledge about the system, which cannot be enforced by traditional input-output system identification methods.}\new{Throughout our experiments, this performance gap increased significantly when 
sparsity patterns or known values of the state-space matrices 
needed to be respected, 
in which case MATLAB often failed to recover meaningful solutions while SIMBa still achieved state-of-the-art accuracy.} Furthermore, on a real-world robot data set, SIMBa often improved state-of-the-art input-state identification methods by more than $70\%$, with the gap widening as more \old{and more }training data became available.

On the other hand, the significant and consistent performance gains observed across the different numerical experiments proposed in this work come with a large computational burden. SIMBa indeed incurred training times ranging from several minutes to two hours in the various case studies. While this might be alleviated in practice by reducing the number of training epochs, augmenting the step size, or initializing SIMBa with matrices known to perform well, it is important to note that the latter does not necessarily improve the final performance.

In \old{the }future \new{works}, it would be interesting to \old{analyze}\new{explore} the theoretical implications of SIMBa\new{, the optimization landscape arising from the different parametrizations,} and \old{explore }potential connections to traditional system identification methods. In another line of work, SIMBa's potential to \new{enforce additional properties on state-space matrices or} be incorporated in Koopman-based approaches with stability guarantees is worth investigating. 

Leveraging the seamless capacity of \texttt{PyTorch} to incorporate various differentiable nonlinear functions, SIMBa has the potential to serve as the foundation for a general tool for knowledge-grounded structured nonlinear system identification.

\section*{\new{Acknowledgments}}

\new{The authors would like to thank Roland T\'oth for his valuable feedback on Sections~\ref{sec:res structure} and~\ref{sec:res naive}.}

{\appendix

    \subsection{Proof of Corollary~\ref{cor:generic}}
    \label{app:corgen}

Let $A$ be a Schur matrix and $\epsilon>0$. \old{Setting $\gamma=1$ in \cite[Theorem 2.2]{chilali1996h}, we know there exists a symmetric Lyapunov function $Q=Q^\top \succ 0$ with 
\begin{equation*}
    \begin{bmatrix}
Q & AQ \\ Q^\top A^\top & Q
    \end{bmatrix} \succ 0 
    \iff  Q - AQ A^\top \succ 0 \;,
\end{equation*}
where the equivalence follows from Schur's complement. }
\old{Introducing a free parameter $G\in\mathbb{R}^{n \times n}$ and the transformation $T = [I, -A]$, this can be rewritten as
\begin{align}
    & Q -AGA^\top -A^\top G^\top A \nonumber\\
    &\qquad +AGA^\top +A^\top G^\top A  - A Q A^\top \succ 0 \nonumber\\
    \iff &T \begin{bmatrix}
         Q & AG \\ G^\top A^\top & G^\top + G - Q
    \end{bmatrix} T^\top \succ 0 \nonumber\\
    \iff & \begin{bmatrix}
        Q & AG \\ G^\top A^\top & G^\top + G - Q
    \end{bmatrix} =: \Gamma \succ 0 \;. \label{eq:generic LMI}
\end{align}}
\new{According to \cite[Theorem 1]{de1999new} we know there exist a symmetric matrix $Q \succ 0$ and $G\in\mathbb{R}^{n \times n}$ such that}
\begin{align}
     \begin{bmatrix}
        Q & AG \\ G^\top A^\top & G^\top + G - Q
    \end{bmatrix} =: \Gamma \succ 0 \;. \label{eq:generic LMI}
\end{align}
In words, if $A$ is Schur stable, there exists $Q \succ 0$ and $G$ such that~\eqref{eq:generic LMI} holds. 
Then, for any $\alpha>0$, $\alpha Q$ and $\alpha G$ are valid alternative choices of Lyapunov function and free parameter because $\alpha Q = \alpha Q^\top \succ 0$ and
\begin{equation*}
    \begin{bmatrix}
    \alpha Q & A (\alpha G) \\ \alpha G^\top A^\top & \alpha G^\top + \alpha G - \alpha Q
\end{bmatrix} = \alpha \Gamma \succ 0 \;.
\end{equation*}
Since $\Gamma$ is positive definite, $\lambda_\textit{min}(\Gamma)>0$, and we can set
\begin{equation*}
    \alpha:=\frac{2\epsilon}{\lambda_\textit{min}(\Gamma)}>0\;, \qquad \Delta := \alpha \Gamma - \epsilon \mathbb{I}_{2n}\;.
\end{equation*}
Then, according to Weyl's inequality~\cite{horn2012matrix}, we have
\begin{align*}
    \lambda_\textit{min}(\Delta) &\geq \lambda_\textit{min}(\alpha \Gamma) - \lambda_\textit{max}(\epsilon \mathbb{I}_{2n}) \\
    &= \alpha \lambda_\textit{min}(\Gamma) - \epsilon = 2\epsilon - \epsilon =  \epsilon > 0\;,
\end{align*}
so that $\Delta = \Delta^\top \succ 0$. We can then define $W=\Delta^{\frac{1}{2}}$, set $V=0$ and construct $A$ as in~\eqref{eq:A generic}, with $S$ as in~\eqref{eq:S generic}.  \hfill $\square$
    \subsection{Proof of Proposition~\ref{prop:continuous}}
    \label{app:continuous}
    
We need to show that $A$ as defined in~\eqref{eq:A generic} is Schur, \oldre{i.e., the autonomous system \begin{equation}
    x_{k+1} = Ax_k \label{eq:autonomous}
\end{equation} 
is stable. This}\newre{which} is equivalent to finding a matrix  $Q=Q^\top \succ 0$ that solves the following Lyapunov inequality~\cite{de1999new}:
\begin{equation} \label{eq:lyap}
    Q - A^\top Q A \succ 0\;.
\end{equation}
By definition of $A$, we have
\begin{align*}
    Q - A^\top Q A &\succ 0 
    &\iff Q - (\mathbb{I}_n + \delta \bar{A})^\top Q (\mathbb{I}_n + \delta \bar{A}) &\succ 0 \\
    \oldre{\iff Q - Q - \delta Q \bar{A} - \delta \bar{A}^\top Q - \delta^2 \bar{A}^\top Q \bar{A} &\succ 0 \\}
    &&\iff - \delta Q \bar{A} - \delta \bar{A}^\top Q - \delta^2 \bar{A}^\top Q \bar{A} &\succ 0\;.
\end{align*}
Let us decompose $\bar{A}$ as $\bar{A} = E^{-1}F$ for suitable matrices $E$ and $F$. We can then rewrite the last inequality as
\begin{align*}
    - \delta Q E^{-1}F - \delta F^\top E^{-\top} Q - \delta^2 F^\top E^{-\top} Q E^{-1}F &\succ 0\;.
\end{align*}
Defining $P=E^{-\top}QE^{-1} \succ 0$ and dividing by $\delta$, \newre{we have}\oldre{this can be rewritten as}
\begin{align} \label{eq:prop2LMI}
    - Q E^{-1}F - F^\top E^{-\top} Q - \delta F^\top PF &\succ 0\;.
\end{align}
Since $Q=E^\top P E$, \eqref{eq:prop2LMI} is equivalent to
\begin{align*}
    - E^\top P F - F^\top P E - \delta F^\top PF &\succ 0 \;.
\end{align*}
Using Schur's complement, this can be rewritten as
\begin{align}
    \begin{bmatrix}
        - E^\top P F - F^\top P E & F^\top \\ F & \frac{1}{\delta} P^{-1}
    \end{bmatrix} \succ 0 \;. \label{eq:app cont}
\end{align}
In words, $A$ as in~\eqref{eq:A continuous} is Schur if and only if there exist $P \succ 0$, $E$, and $F$ such that $\bar{A} = E^{-1}F$ and~\eqref{eq:generic LMI} holds.

Let us now parametrize the left-hand side of the above LMI with $S$ from~\eqref{eq:S continuous}. Critically, \eqref{eq:app cont} will always be satisfied since $S$ is positive definite by construction for any choice of $W$, ensuring the stability of $A$. Since $S_{22}=S_{22}^\top$, we \oldre{can }then recover
\begin{align*}
    P &= \frac{S_{22}^{-1}}{\delta}\;, & F &= S_{21} = S_{12}^\top\;. 
\end{align*}
Knowing that $S_{11}^\top = S_{11}$ by definition and that
\begin{align*}
    -E^\top P F -F^\top P E &= S_{11} 
\end{align*}
needs to hold, we can set
\begin{align*}
    F^\top P E &= -\frac{S_{11}}{2} + V - V^\top \\
    \implies E^{-1} &= -2 \left(S_{11} + V - V^\top\right)^{-1} F^\top P\;, 
\end{align*}
for any $V\in\mathbb{R}^{n \times n}$. Since $\bar{A}=E^{-1}F$, this leads to
\begin{align*}
    \bar{A} &= - 2 \left(S_{11} + V - V^\top\right)^{-1} F^\top P F \\
    &= - \frac{2}{\delta}\left(S_{11} + V - V^\top\right)^{-1}S_{12}S_{22}^{-1}S_{21}
\end{align*}
and
\begin{align*}
    A &= I + \delta \bar{A} = I - 2 \left(S_{11} + V - V^\top\right)^{-1} S_{12} S_{22}^{-1} S_{21}\;. \hfill \square
\end{align*}

    \subsection{Proof of Corollary~\ref{cor:continuous}}
    \label{app:corcont}
    
Let $A$ be a Schur matrix and $\epsilon>0$. Setting $E=\mathbb{I}_{n}$ and $F=\bar{A}$ in the proof of Proposition~\ref{prop:continuous}, there exists a Lyapunov function $P = P^\top \succ 0$ such that~\eqref{eq:app cont} holds. Similarly to the proof of Corollary~\ref{cor:generic}, for any $\alpha>0$, $\alpha E$, $\alpha F$, and  $\frac{P}{\alpha}$ are valid alternative choices since
\begin{equation*}
    \begin{bmatrix}
    - \alpha E^\top \frac{P}{\alpha} \alpha F - \alpha F^\top \frac{P}{\alpha} \alpha E & \alpha F^\top \\ \alpha F & \frac{1}{\delta} \alpha P^{-1}
\end{bmatrix} \succ 0
\end{equation*}
and $(\alpha E)^{-1} \alpha F = \bar{A}$. One can then set $V=0$ and follow the proof of Corollary~\ref{cor:generic} to find suitable values for $\alpha$ and $W$ such that~\eqref{eq:A continuous} holds for $S$ as in~\eqref{eq:S continuous}. \hfill $\square$

    \subsection{Proof of Proposition~\ref{prop:sparse}}
    \label{app:sparse}

    \subsubsection{Preliminaries}

Throughout the proof below, we will use the following properties of the Hadamard product, for any matrices $K,L,M\in\mathbb{R}^{n \times n}$ and diagonal matrix $\Lambda\in\mathbb{R}^{n \times n}$~\cite{horn2012matrix}:
\begin{itemize}
    \item[(P1)] $ (K \odot L) + (K \odot M) = K \odot (L+M)$\;,
    \item[(P2)] $(L \odot M)^\top = (L^\top \odot M^\top)$\;, 
    \item[(P3)] $ (K \odot L)(\Lambda \odot M) = K \odot \left(L(\Lambda \odot M)\right)$\;, 
\end{itemize}
where the last property holds since $\Lambda$ is diagonal.

    \subsubsection{Proof of Proposition~\ref{prop:sparse}}

As in the proof of Proposition~\ref{prop:continuous}, we need to find a symmetric matrix $Q = Q^\top \succ 0$ such that $Q - A^\top Q A \succ 0$. Following the proof of Corollary~\ref{cor:generic}, one can show that this is equivalent to finding $Q=Q^\top \succ 0$ such that
\begin{align}
    \begin{bmatrix}
        Q & AG \\ G^\top A^\top & G^\top + G - Q
    \end{bmatrix} \succ 0\;, \label{eq:sparse LMI}
\end{align}
\new{for some}\old{where} $G\in\mathbb{R}^{n \times n}$\old{ is a free parameter}.
In this sparse case, we consider diagonal $Q$ and $G$ matrices of the form 
\begin{align*}
    Q &= N \odot P\;, &G &= N \odot H\;,
\end{align*}
for some $P, H\in\mathbb{R}^{n \times n}$ and with $N$ defined in~\eqref{eq:N sparse}. Note here that this also implies $Q=Q^\top$, as required in~\eqref{eq:sparse LMI}. Recalling that we want to identify a matrix $A$ of the form $A = \mathcal{M} \odot \tilde{A}$ for some $\tilde{A}$, \eqref{eq:sparse LMI} can be written as
\begin{align*}
    \begin{bmatrix}
        N \odot P & \left(\mathcal{M} \odot \tilde{A}\right) \left(N \odot H\right) \\ (*)^\top & \left(N \odot H\right)^\top + \left(N \odot H\right) - \left(N \odot P\right)
    \end{bmatrix} &\succ 0  \\
    \overset{\text{(P1)}}{\iff} \begin{bmatrix}
        N \odot P & \left(\mathcal{M} \odot \tilde{A}\right) \left(N \odot H\right) \\ (*)^\top & N \odot \left(H^\top + H - P\right)
    \end{bmatrix} &\succ 0\;, 
\end{align*}
where $(*)^\top$ represents the transpose of the upper right block. Since $N$ is a diagonal matrix, using (P3), we can rewrite the above LMI as 
\begin{align}
    \begin{bmatrix}
        N \odot P & \mathcal{M} \odot (\tilde{A} (N \odot H)) \\ ( \mathcal{M} \odot (\tilde{A} (N \odot H)))^\top & N \odot \left(H^\top + H - P\right)
    \end{bmatrix} &\succ 0 \nonumber \\ 
    \overset{\text{(P2)}}{\iff} \begin{bmatrix}
        N \odot P & \mathcal{M} \odot (\tilde{A} (N \odot H)) \\ \mathcal{M}^\top \odot (\tilde{A} (N \odot H))^\top  & N \odot \left(H^\top + H - P\right)
    \end{bmatrix} &\succ 0 \nonumber \\ 
    \iff \begin{bmatrix}
        N & \mathcal{M} \\ (*)^\top & N 
    \end{bmatrix} \odot 
    \begin{bmatrix}
        P & \tilde{A} (N \odot H) \\ (*)^\top & H^\top + H - P
    \end{bmatrix} &\succ 0\;. \label{eq:app sparse}
\end{align}
A sufficient condition for the above Hadamard product to be positive semi-definite is to ensure that both factors are individually positive semi-definite~\cite{zhang2006schur}, i.e.,
\begin{align}
    &\begin{bmatrix}
        N & \mathcal{M} \\ (*)^\top & N 
    \end{bmatrix} \succ 0 \label{eq:levy} \\
    &\begin{bmatrix}
        P & \tilde{A} (N \odot H) \\ (*)^\top & H^\top + H - P
    \end{bmatrix} \succ 0\;. \label{eq:sparse LMI2}
\end{align}

Since $\mathcal{M}$ is fixed and known, \eqref{eq:levy} is satisfied by construction of $N$ in~\eqref{eq:N sparse} according to the 
Levy–Desplanques theorem~\cite{horn2012matrix}. To satisfy~\eqref{eq:sparse LMI2}, as in Proposition~\ref{prop:continuous}, we parametrize its left-hand side with $S$ in~\eqref{eq:S sparse}, which allows us to recover
\begin{align*}
    P &= S_{11}\;, &H + H^\top = S_{22} + P\;.
\end{align*}
As before, by symmetry of $S_{11}$ and $S_{22}$, for any $V$, the right equation is satisfied for
\begin{equation*}
    H = \frac{1}{2} (S_{11} + S_{22}) + V - V^\top\;.
\end{equation*}
\oldre{Finally, 
\begin{align*}
    \tilde{A} &= S_{12}(N \odot H)^{-1} \\
    &= S_{12}\left[N \odot \left(\frac{1}{2} (S_{11} + S_{22}) + V - V^\top\right)\right]^{-1}
\end{align*}
and
\begin{align*}
    A &= \mathcal{M} \odot \tilde{A} \\
    &= \mathcal{M} \odot \left(S_{12}\left[N \odot \left(\frac{1}{2} (S_{11} + S_{22}) + V - V^\top\right)\right]^{-1}\right). \hfill \square
\end{align*}}
\newre{
Finally, 
\begin{align*}
    A &= \mathcal{M} \odot \tilde{A} = \mathcal{M} \odot \left( S_{12}(N \odot H)^{-1} \right) \\
    &= \mathcal{M} \odot \left(S_{12}\left[N \odot \left(\frac{1}{2} (S_{11} + S_{22}) + V - V^\top\right)\right]^{-1}\right). \hfill \square
\end{align*}
}

    \subsection{Proof of Proposition~\ref{prop:naive}}
    \label{app:naive}
 
First, the desired sparsity pattern is achieved because 
\begin{equation*}
    \frac{\sigma(\eta)\gamma}{|\lambda(\mathcal{M} \odot V)|_{\textit{max}}} \left(\mathcal{M} \odot V\right) = \mathcal{M} \odot  \left(\frac{\sigma(\eta)\gamma}{|\lambda(\mathcal{M} \odot V)|_{\textit{max}}}V\right)
\end{equation*} 
by definition of the Hadamard product since the maximum eigenvalue is a scalar. 

To show that $A$ is Schur with eigenvalues in a circle of radius $\gamma$ centered at the origin, suppose $\beta$ is an eigenvalue of $\left(\mathcal{M} \odot V\right)$ corresponding to the eigenvector $e$, i.e., $\left(\mathcal{M} \odot V\right)e = \beta e$. Then, 
\begin{align*}
    Ae &= \frac{\sigma(\eta)\gamma}{|\lambda(\mathcal{M} \odot V)|_{\textit{max}}} \left(\mathcal{M} \odot V\right)e 
    =: \alpha e \;,
\end{align*}

so that $e$ is still an eigenvector of $A$, with eigenvalue $\alpha$. By definition of $|\lambda(\mathcal{M} \odot V)|_{\textit{max}}$\old{ and}\new{,} since $0 < \sigma(\eta) < 1$, we obtain
\begin{equation*}
    |\alpha| = \frac{\sigma(\eta)\gamma}{|\lambda(\mathcal{M} \odot V|_{\textit{max}}} |\beta| < \frac{|\beta|}{|\lambda(\mathcal{M} \odot V)|_{\textit{max}}} \gamma \leq \gamma\;,
\end{equation*}
hence, concluding the proof. \hfill $\square$

\bibliographystyle{IEEEtran}
\bibliography{references}

\begin{IEEEbiography}[{\includegraphics[width=1in,height=1.25in,clip,keepaspectratio]{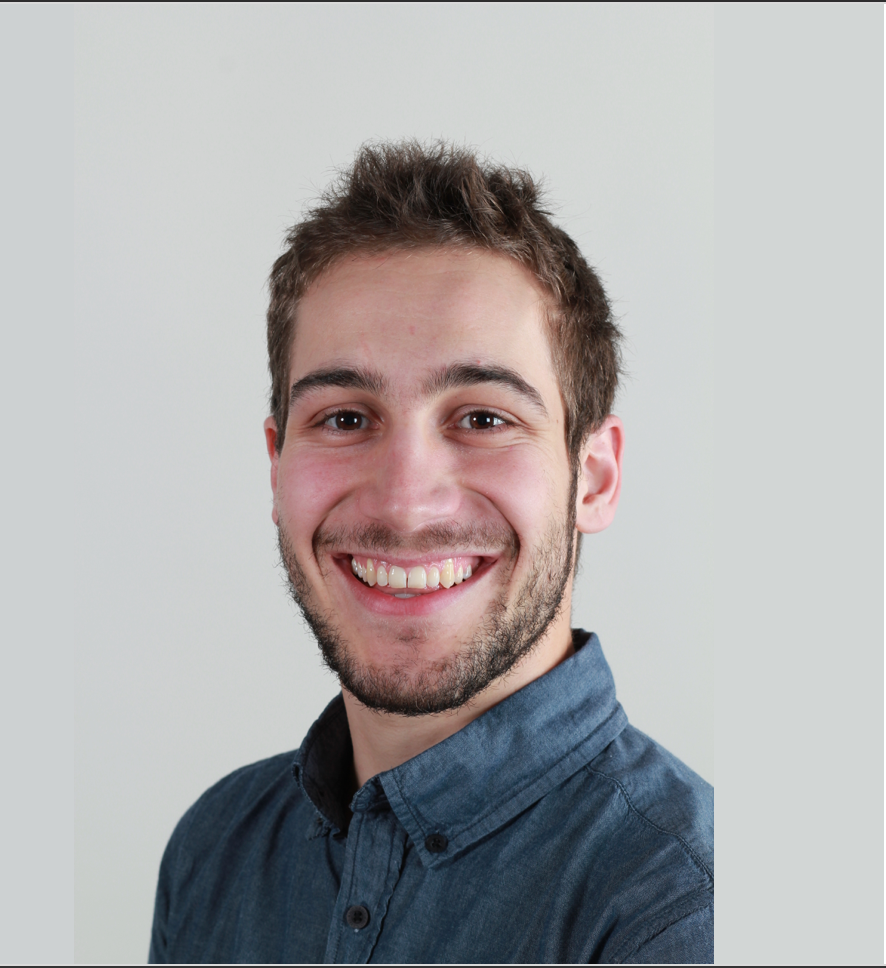}}]{Loris Di Natale} 
received his B.Sc. in Mathematics in 2017 and his M.Sc. in Energy Management and Sustainability in 2019, both from EPFL. He received his Ph.D. in Electrical Engineering from EPFL in 2024, where he was jointly working at the Urban Energy Systems Lab at Empa and the Automatique Control Lab at EPFL, both in Switzerland. He his currently working as an AI and control engineer for Zürich Soft Robotics GmbH. His research interests encompass Machine Learning and its applications to Energy Systems.
\end{IEEEbiography}
\begin{IEEEbiography}[{\includegraphics[width=1in,height=1.25in,clip,keepaspectratio]{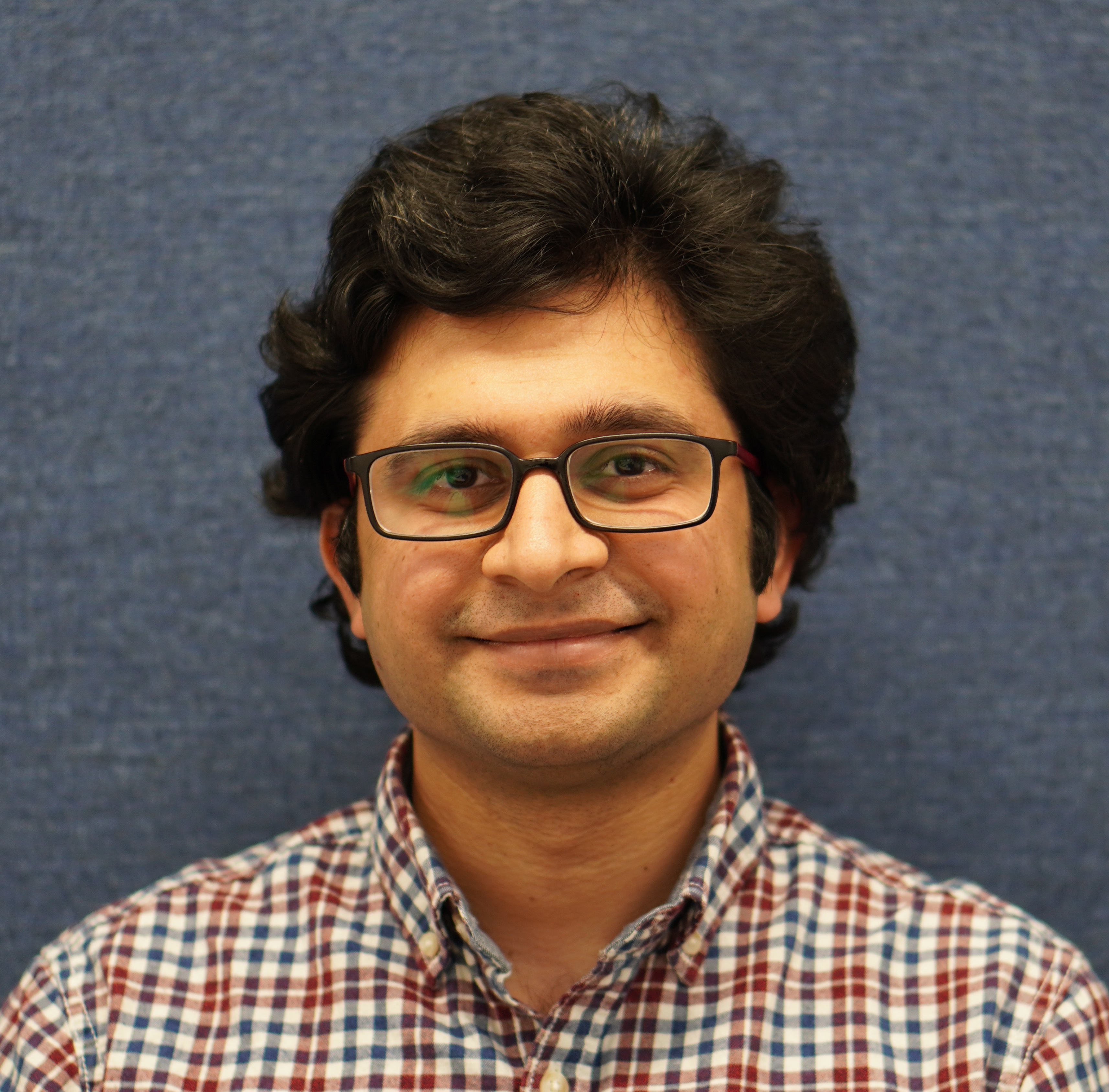}}]{Muhammad Zakwan}
 is a doctoral assistant in the Dependable Control and Decision group (DECODE) at École Polytechnique Fédérale de Lausanne (EPFL). He is also a member of the National Centre of Competence in Research (NCCR) Automation. He holds an Electrical and Electronics  Engineering degree from Bilkent University (Turkey) and a Bachelor of Science in Electrical Engineering from the Pakistan Institute of Engineering and Applied Sciences (Pakistan). His research interests include neural networks, nonlinear control, contraction theory,  port-Hamiltonian systems, and machine learning.
\end{IEEEbiography}
\begin{IEEEbiography}[{\includegraphics[width=1in,height=1.25in,clip,keepaspectratio]{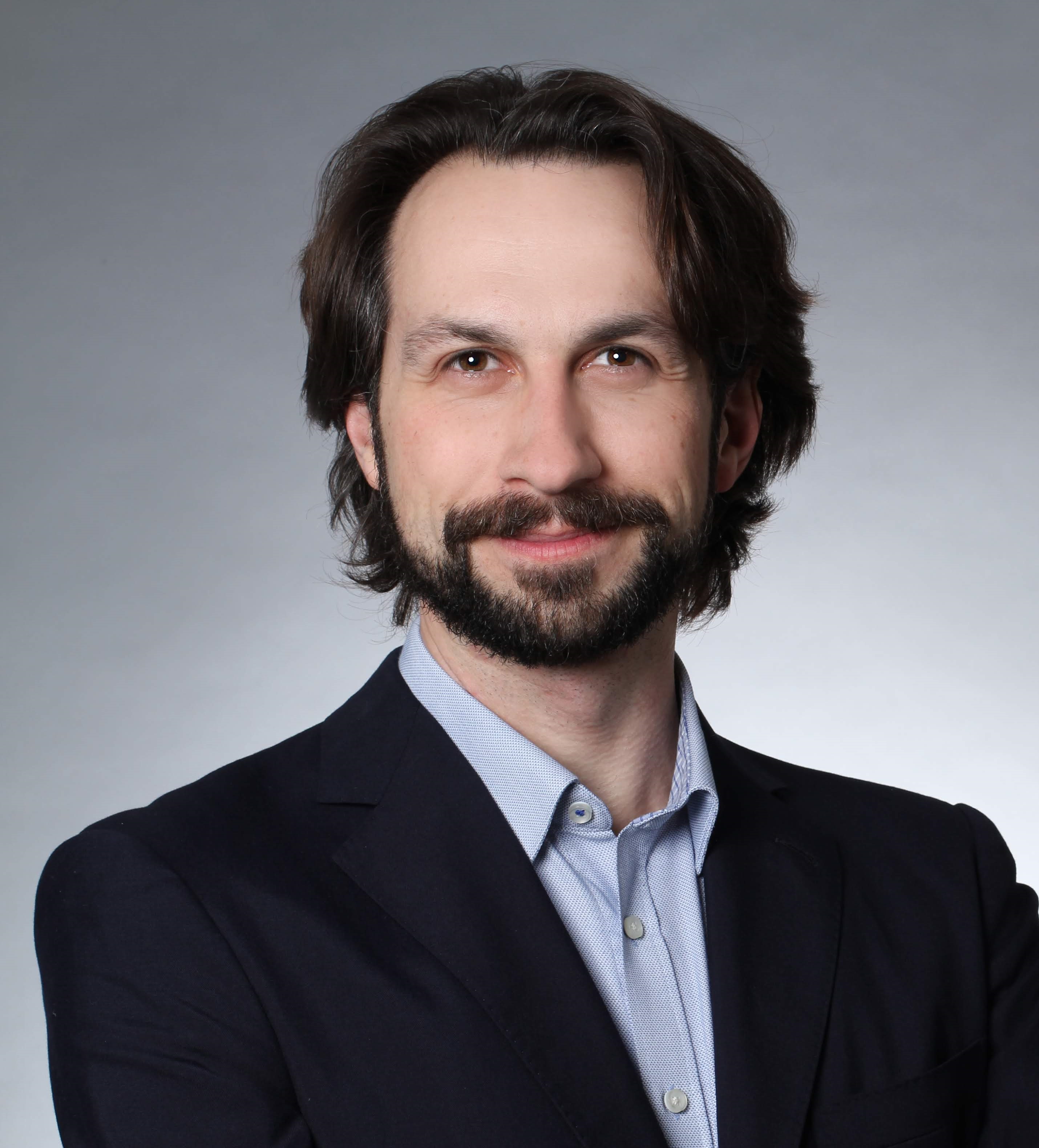}}]{Philipp Heer}
is the deputy head of the Urban Energy Systems Lab (UESL) at Empa - an Institute of the ETH Domain – and is responsible for the energy and digitalization related research at the Empa demonstrators NEST, move and ehub. He received a B.Sc. from HSLU T\&A (2010) as well as a M.Sc. from ETH Zurich (2013) both in Electrical, Electronics and Communications Engineering and a MAS from ETH Zurich (2018) in Management, Technology, and Economics. His research focuses on developing operational solutions based on data-driven approaches to foster sustainable local energy systems. 
\end{IEEEbiography}
\begin{IEEEbiography}[{\includegraphics[width=1in,height=1.25in,clip,keepaspectratio]{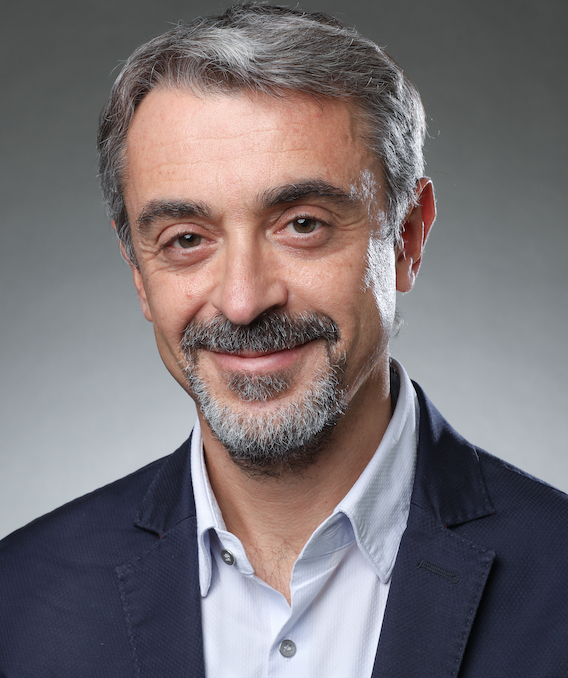}}]{Giancarlo Ferrari Trecate}
(SM'12) received a Ph.D. in Electronic and Computer Engineering from the Universita' Degli Studi di Pavia in 1999. Since September 2016, he has been a Professor at EPFL, Lausanne, Switzerland. In the spring of 1998, he was a Visiting Researcher at the Neural Computing Research Group, University of Birmingham, UK. In the fall of 1998, he joined the Automatic Control Laboratory, ETH, Zurich, Switzerland, as a Postdoctoral Fellow. He was appointed Oberassistent at ETH in 2000. In 2002, he joined INRIA, Rocquencourt, France, as a Research Fellow. From March to October 2005, he was a researcher at the Politecnico di Milano, Italy. From 2005 to August 2016, he was Associate Professor at the Dipartimento di Ingegneria Industriale e dell'Informazione of the Universita' degli Studi di Pavia.
His research interests include scalable control, machine learning, microgrids, networked control systems, and hybrid systems.
Giancarlo Ferrari Trecate is the founder and current chair of the Swiss chapter of the IEEE Control Systems Society. He is Senior Editor of the IEEE Transactions on Control Systems Technology and has served on the editorial boards of Automatica and Nonlinear Analysis: Hybrid Systems.
\end{IEEEbiography}
\begin{IEEEbiography}[{\includegraphics[width=1in,height=1.25in,clip,keepaspectratio]{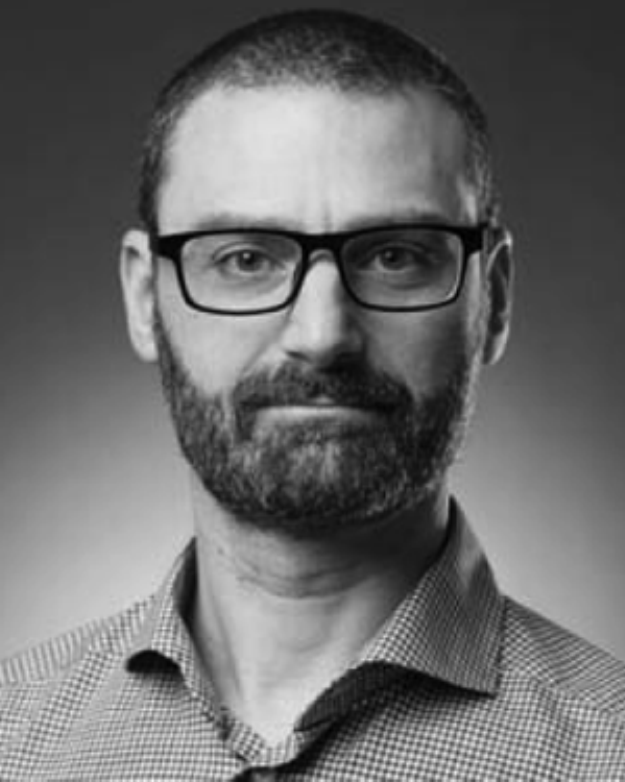}}]{Colin N. Jones}
(Senior Member) received the bachelor’s and master’s degrees in electrical engineering from The University of British Columbia, Vancouver, BC, Canada, in 1999 and 2001, respectively, and the Ph.D. degree, for his work on polyhedral computational methods for constrained control, from the University of Cambridge, Cambridge, U.K., in 2005. He was a Senior Researcher with the Automatic Control Laboratory, ETH Zürich, Switzerland, until 2010 and an Assistant and then Associate Professor with the Automatic Control Laboratory, EPFL, Switzerland, since 2011. His current research interests include high-speed predictive control and optimization, and the control of green energy generation, distribution, and management. He serves as an associate editor for the Transactions on Automatic Control, Control Systems Letters and the Transactions on Control Systems Technology. Dr. Jones has received the ERC Starting Grant to study the optimal control of building networks and is the author or coauthor of more than 200 publications.
\end{IEEEbiography}

\vfill

\end{document}